\documentclass[twocolumn]{aastex631}

\newcommand{\civ}{C~{\sc iv}}
\newcommand{\Siiv}{Si~{\sc iv}}
\newcommand{\heii}{He~{\sc ii}}
\newcommand{\ciii}{C~{\sc iii}]}
\newcommand{\hb}{H$\beta$}
\newcommand{\ha}{H$\alpha$}
\newcommand{\hg}{H$\gamma$}
\newcommand{\mgii}{Mg~{\sc ii}}
\newcommand{\feii}{Fe~{\sc ii}}

\newcommand{\swift}{SWIFT}
\newcommand{\Lbol}{$L_{\rm bol}$}
\newcommand{\MBH}{M$_{\rm BH}$}
\newcommand{\Msun}{M$_{\odot}$}
\newcommand{\cosstis}{{\it COS-STIS}}
\newcommand{\ergs}{erg\,s$^{-1}$}
\newcommand{\Ledd}{$L_{\rm edd}$}
\newcommand{\kms}{km~s$^{-1}$}
\newcommand{\lya}{Ly-$\alpha$}
\newcommand{\Ldisk}{$L_{\rm disk}$}
\newcommand{\Rg}{$R_{\rm g}$}
\newcommand{\cc}{cm$^{-3}$}

\received{August 2, 2024}
\revised{September 20, 2024}
\accepted{September 30, 2024}

\begin{document}

\title[interband continuum delays in Mrk~817 ]{AGN STORM 2: X. The origin of the interband continuum delays in Mrk~817$^*$}\footnote{$*$Based in part on observations associated with program GO-16196 made with the NASA/ESA Hubble Space Telescope, obtained at the Space Telescope Science Institute, which is operated by the Association of Universities for Research in Astronomy, Inc., under NASA contract NAS5-26555.}


\author[0000-0002-6766-0260]{Hagai Netzer}
\affiliation{School of Physics and Astronomy, Tel Aviv University, Tel Aviv 6997801, Israel}

\author[0000-0002-2908-7360]{Michael R.\ Goad}
\affiliation{School of Physics and Astronomy, University of Leicester, University Road, Leicester, LE1 7RH, UK}

\author[0000-0002-3026-0562]{Aaron J.\ Barth}
\affiliation{Department of Physics and Astronomy, 4129 Frederick Reines Hall University of California, Irvine, CA, 92697-4575, USA}

\author[0000-0002-8294-9281]{Edward M.\ Cackett}
\affiliation{Department of Physics and Astronomy, Wayne State University, 666 W.\ Hancock St, Detroit, MI, 48201, USA}

\author[0000-0003-1728-0304]{Keith Horne}
\affiliation{SUPA School of Physics and Astronomy, North Haugh, St.~Andrews, KY16~9SS, Scotland, UK}

\author{Chen Hu}
\affiliation{Key Laboratory for Particle Astrophysics, Institute of High Energy Physics, Chinese Academy of Sciences, 19B Yuquan Road,\\ Beijing 100049, People's Republic of China}

\author[0000-0003-0172-0854]{Erin Kara}
\affiliation{MIT Kavli Institute for Astrophysics and Space Research, Massachusetts Institute of Technology, Cambridge, MA 02139, USA}

\author[0000-0003-0944-1008]{Kirk T.\ Korista}
\affiliation{Department of Physics, Western Michigan University, 1120 Everett Tower, Kalamazoo, MI 49008-5252, USA}

\author[0000-0002-2180-8266]{Gerard A.\ Kriss}
\affiliation{Space Telescope Science Institute, 3700 San Martin Drive, Baltimore, MD 21218, USA}

\author[0000-0002-8671-1190]{Collin Lewin}
\affiliation{MIT Kavli Institute for Astrophysics and Space Research, Massachusetts Institute of Technology, Cambridge, MA 02139, USA}

\author[0000-0001-5639-5484]{John Montano}
\affiliation{Department of Physics and Astronomy, 4129 Frederick Reines Hall, University of California, Irvine, CA, 92697-4575, USA}


\author[0000-0003-2991-4618]{Nahum Arav}
\affiliation{Department of Physics, Virginia Tech, Blacksburg, VA 24061, USA}





\author[0000-0001-9735-4873]{Ehud Behar}
\affiliation{Department of Physics, Technion, Haifa 32000, Israel}
\affiliation{MIT Kavli Institute for Astrophysics and Space Research, Massachusetts Institute of Technology, Cambridge, MA 02139, USA}

\author[0000-0002-1207-0909]{Michael S.\ Brotherton}
\affiliation{Department of Physics and Astronomy, University of Wyoming, Laramie, WY 82071, USA}

\author{Doron Chelouche}
\affiliation{Department of Physics, Faculty of Natural Sciences, University of Haifa, Haifa 3498838, Israel}
\affiliation{Haifa Research Center for Theoretical Physics and Astrophysics, University of Haifa, Haifa 3498838, Israel}

\author[0000-0003-3242-7052]{Gisella De~Rosa}
\affiliation{Space Telescope Science Institute, 3700 San Martin Drive, Baltimore, MD 21218, USA}

\author[0000-0001-9931-8681]{Elena Dalla Bont\`{a}}
\affiliation{Dipartimento di Fisica e Astronomia ``G.\  Galilei,'' Universit\`{a} di Padova, Vicolo dell'Osservatorio 3, I-35122 Padova, Italy}
\affiliation{INAF-Osservatorio Astronomico di Padova, Vicolo dell'Osservatorio 5 I-35122, Padova, Italy}
\affiliation{Jeremiah Horrocks Institute, University of Central Lancashire, Preston, PR1 2HE, UK}

\author[0000-0002-0964-7500]{Maryam Dehghanian}
\affiliation{Department of Physics, Virginia Tech, Blacksburg, VA 24061, USA}

\author[0000-0003-4503-6333]{Gary J.\ Ferland}
\affiliation{Department of Physics and Astronomy, The University of Kentucky, Lexington, KY 40506, USA}


\author[0000-0002-2306-9372]{Carina Fian}
\affiliation{Departamento de Astronom\'{i}a y Astrof\'{i}sica, Universidad de Valencia, E-46100 Burjassot, Valencia, Spain}
\affiliation{ Observatorio Astron\'{o}mico, Universidad de Valencia, E-46980 Paterna, Valencia, Spain}






\author[0000-0002-0957-7151]{Yasaman Homayouni}
\affiliation{Space Telescope Science Institute, 3700 San Martin Drive, Baltimore, MD 21218, USA}
\affiliation{Department of Astronomy and Astrophysics, The Pennsylvania State University, 525 Davey Laboratory, University Park, PA 16802}
\affiliation{Institute for Gravitation and the Cosmos, The Pennsylvania State University, University Park, PA 16802}



\author[0000-0002-1134-4015]{Dragana Ili\'{c}}
\affiliation{University of Belgrade - Faculty of Mathematics, Department of Astronomy, Studentski trg 16, 11000 Belgrade, Serbia}
\affiliation{Hamburger Sternwarte, Universit{\"a}t Hamburg, Gojenbergsweg 112, 21029 Hamburg, Germany}


i

\author[0000-0002-9925-534X]{Shai Kaspi}
\affiliation{School of Physics and Astronomy and Wise observatory, Tel Aviv University, Tel Aviv 6997801, Israel}



\author[0000-0001-5139-1978]{Andjelka B. Kova{\v c}evi{\'c}}
\affiliation{University of Belgrade - Faculty of Mathematics, Department of Astronomy, Studentski trg 16, 11000 Belgrade, Serbia}







\author{Hermine Landt}
\affiliation{Centre for Extragalactic Astronomy, Department of Physics, Durham University, South Road, Durham DH1 3LE, UK}




\author[0000-0003-2398-7664]{Luka \v{C}.\ Popovi\'{c}}
\affiliation{Astronomical Observatory, Volgina 7, 11060 Belgrade, Serbia}
\affiliation{University of Belgrade - Faculty of Mathematics, Department of Astronomy, Studentski trg 16, 11000 Belgrade, Serbia}


\author[0000-0003-1772-0023]{Thaisa Storchi-Bergmann}
\affiliation{Departamento de Astronomia - IF, Universidade Federal do Rio Grande do Sul, CP 150501, 91501-970 Porto Alegre, RS, Brazil}





\author[0000-0001-9449-9268]{Jian-Min Wang}
\affiliation{Key Laboratory for Particle Astrophysics, Institute of High Energy Physics, Chinese Academy of Sciences, 19B Yuquan Road,\\ Beijing 100049, People's Republic of China}
\affiliation{School of Astronomy and Space Sciences, University of Chinese Academy of Sciences, 19A Yuquan Road, Beijing 100049, People's Republic of China}
\affiliation{National Astronomical Observatories of China, 20A Datun Road, Beijing 100020, People's Republic of China}





\author[0000-0003-0931-0868]{Fatima Zaidouni}
\affiliation{MIT Kavli Institute for Astrophysics and Space Research, Massachusetts Institute of Technology, Cambridge, MA 02139, USA}

\begin{abstract}

The local (z=0.0315) AGN Mrk~817, was monitored over more than 500
days with space-borne and ground-based instruments as part of a large
international campaign AGN STORM~2.  Here, we present a comprehensive
analysis of the broad-band continuum variations using detailed
modeling of the broad line region (BLR), several types of disk winds
classified by their optical depth, and new numerical simulations.
We find that diffuse continuum (DC) emission, with additional
contributions from strong and broad emission lines, can explain the
continuum lags observed in this source during high and low luminosity
phases.  Disk illumination by the variable X-ray corona contributes
only a small fraction of the observed continuum lags.
  Our BLR models assume radiation pressure-confined clouds distributed
  over a distance of 2-122 light days.  We present calculated
  mean-emissivity radii of many emission lines, and DC emission, and
  suggest a simple, transfer-function-dependent method that ties them
  to cross-correlation lag determinations.
We do not find clear indications for large optical depth winds but
identify the signature of lower column density winds. In particular,
we associate the shortest observed continuum lags with a combination
of $\tau$(1\,Ryd)$\approx 2$ wind and a partly shielded BLR.  Even
smaller optical depth winds may be associated with X-ray absorption
features and with noticeable variations in the width and lags of
several high ionization lines like \heii\ and \civ.  Finally, we
demonstrate the effect of torus dust emission on the observed lags in
the $i$ and $z$ bands.

\end{abstract}

%


\section{Introduction}
Active galactic nuclei (AGN) are known for their very high luminosity
originating from the vicinity of the supermassive black hole (SMBH) in
their center and for their highly variable continuum and emission
lines. Two types of studies aimed at understanding the nature of these
variations have been carried out over several decades. One type
involves high-cadence monitoring of a large number of AGN in a small
number of UV and optical bands \citep[see,
  e.g.,][]{Kaspi2000,Bentz2013, Shen2016,Lira2018,Du2019}. Such
reverberation mapping (RM) campaigns reveal a strong correlation
between the variation of the hydrogen Balmer lines and the optical
continuum, where the line flux follows the continuum flux with
luminosity and Eddington ratio-dependent lags. Similar
luminosity-dependent correlations between \civ\ and other strong
emission lines and the UV/optical continuum have also been found. Such
correlations have become a standard tool for measuring the BH mass in
AGN at all redshifts and luminosities.

A second type of campaign is long-term multi-wavelength monitoring of
individual sources to find luminosity-dependent lags between different
bands or between emission lines and continuum bands. Some studies
focus on broad-band observations, looking for a luminosity-dependent
correlation between X-ray and UV or X-ray and optical variations
(e.g., the large Fairall-9 campaign described in
\cite{Hernandez2020}). While statistically significant correlations
between the X-ray and optical continua are hard to find, strong
correlations between several optical and UV continuum bands are
widespread. Among the most successful long-term spectroscopic studies,
we mention the 2014 AGN STORM~1 campaign, which aimed to monitor the
X-ray to infrared continuum and emission-line variations in NGC~5548
\citep{deRosa2015, Fausnaugh2016, Pei2017}.

AGN STORM~2 is a large monitoring campaign that follows a local
(z=0.0315) AGN, Mrk~817, over a period that has already exceeded 500
days. The project combines spectroscopic and photometric data in
several wavelength bands, from infrared to hard X-rays. Telescopes and
instruments involved are the Hubble Space Telescope (HST) COS and STIS
spectrographs, \swift, XMM, NuStar, NICER, and several small-to-medium
size ground-based telescopes. The observations show large amplitude
X-ray and UV variations and smaller variations at longer
wavelengths. So far, the project has resulted in 9 published and
submitted papers: \citet{Kara2021,
  Homayouni2023,Partington2023,Cackett2023,Homayouni2024,Neustadt2024,Dehghanian2024,Zaidouni2024},
and Lewin et al. (2024, hereafter L24).

Spectroscopic RM analysis based on AGN STORM~2 published and
unpublished data (C. Hu, private communication) suggests a lag of
about 25$\pm 5$ days between the 5100\AA\ continuum and the \hb\ line,
depending on the exact part of the continuum light curve used in the
analysis. There are also measured lags between the \civ\ emission line
and the UV continuum, which change between 3 and 12 days throughout
the campaign \cite[see][]{Homayouni2024}.  Broad-band photometry,
combining HST, \swift, and ground-based facilities, suggest lags
relative to the HST/1144\AA\ rest wavelength continuum that increase
with wavelength from about two days for the 2000-4000\AA\ continuum to
more than four days in the $i$ and $z$ bands (see \cite{Cackett2023},
hereafter C23, and L24). The X-ray to UV variations are hard to
interpret, and no clear lags persisting over long periods have been
found. Some papers address temperature variations across the accretion
disk \citep{Neustadt2024}. Still, it is not yet clear what the
connection, if any, is between this study and the total size and
geometry of the disk.

The present paper aims to use detailed photoionization modeling of
radiation-pressure-confined (RPC) clouds
\citep{Baskin2014,Stern2014,Baskin2018, Netzer2020} in the broad line
region (BLR) of Mrk~817 to test the suggestion that the time variable
diffuse emission from this region is the primary source of the
observed optical/UV broad-band lags. Models of this type have already
been applied to other local AGN \citep[e.g.,][]{Netzer2022}. However,
previous attempts did not investigate the physics of the BLR gas and
the properties of the disk wind observed in this source at such a
level of detail. In particular, we try to explain the observed
broad-band line and continuum variations by combining time-dependent
emission from an X-ray illuminated disk with the variable diffuse
continuum (DC) flux from the BLR. We use the models to address the
size of the accretion disk and the geometry of the BLR in Mrk~817.
 
 Section\,2 of the paper provides information about the variability
 observed in this source. In Section\,3, we present our combined
 disk-BLR model, and in Section\,4 we present new numerical
 simulations that enable us to predict the expected continuum lags
 using a simple, easy-to-calculate expression. The results are shown
 and discussed in Section\,5, where we also present a comprehensive
 analysis of various types of disk winds (DWs) observed in this
 source.  Additional minor but important issues are discussed in the
 appendix.  Throughout the paper, we assume a luminosity distance
 $d_{\rm L}=134.2$~Mpc for Mrk~817 (see
 http://www.astro.gsu.edu/AGNmass/), and cosmological parameters
 $H_{0}=$72~km~s$^{-1}$~Mpc$^{-1}$ \, , $\Omega_{\Lambda}=0.7$, and
 $\Omega_{M}=0.3$.

\section{Observations and basic assumptions}
\label{basic_assumptions}
\subsection{Continuum spectral energy distribution (SED)}

We used the multi-wavelength observations of Mrk~817 described in
earlier papers of AGN STORM 2 to construct a combined X-ray disk-BLR
model for this source. The model follows the computations of
\cite{Slone2012}. It assumes an optically thick geometrically-thin
accretion disk around a BH of mass \MBH=$4 \times 10^7$\Msun\ and spin
parameter {\it a}=0.7, which corresponds to a mass conversion
efficiency of $\eta = 0.1$. Full relativistic corrections and
Comptonization are included in the calculations. The disk spectrum was
combined with a power-law $\Gamma=1.9$ X-ray source normalized to the
disk 2500\AA\ luminosity such that $\alpha_{OX}=-1.45$. The
combination of the two assumes that the source of the emitted X-ray
radiation is disk accretion. Thus the X-ray luminosity is part of the
bolometric luminosity of the system. Given the small contribution of
the X-ray luminosity to \Lbol\ (about 10\%) the angular dependence of
the X-ray radiation is not very important, and complete isotropy was
assumed.

The present SED is normalized differently from the one presented by
\cite{Kara2021}. It considers the disk inclination to the line of
sight, {\it i}, which is chosen to be $\cos i=0.75$. This issue was
discussed in earlier campaign papers \citep[e.g.,][]{Zaidouni2024},
and inclination angles as small as 20 degrees have been
suggested. Here we assume a standard limb-darkening law of $L_{\nu}
\propto \cos i(1+b \cos i)$ with $b=2$ \citep[e.g.,][]{Netzer2013} to
obtain the bolometric luminosity of the disk.

\begin{figure} \centering
\includegraphics[width=0.95\linewidth]{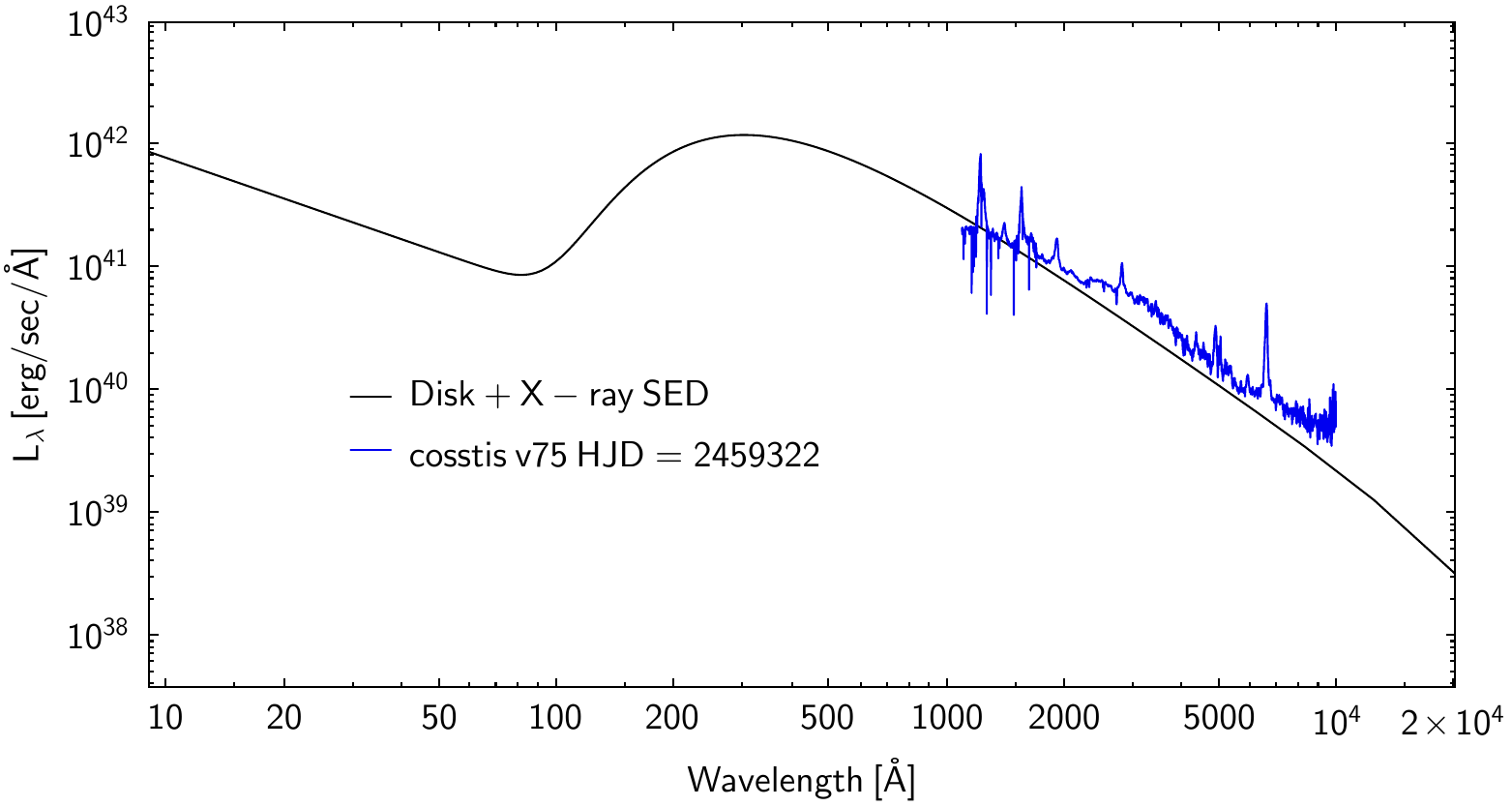}
	\caption{Assumed continuum SED (in black) for Mrk~817
          constructed from a thin disk and a power-law X-ray source
          normalized to the luminosity during the first 100 days of
          the campaign. The disk inclination to the line-of-sight is
          given by $\cos i=0.75$. The blue line is a
          \cosstis\ spectrum obtained on the HJD marked in the
          figure. More details are provided in the text.  }
\label{full_SED}
\end{figure}

A major assumption of the models described below is a large DC
contribution to all wavelengths between 1800\AA\, and 10000\AA. The
luminosity normalization is based on a combined HST/{\it COS-STIS}
(hereafter \cosstis) spectrum obtained on HJD=2459322. This spectrum
is similar to the \cosstis\ spectrum obtained on HJD=2459202, shown
and discussed in \cite{Kara2021}. We assume no contribution from the
host galaxy to this spectrum at all observed wavelengths because of
the narrow entrance aperture (0.1 arcsec). Given all this, the models
in \cite{Netzer2022} as well as new models presented in this paper,
the normalized disk spectrum at the beginning of the campaign is taken
to be $\lambda L_{\lambda}(5100\AA)=5.3 \times 10^{43}$~\ergs, and the
DC contribution at 5100\AA\ is about 20-30\%.  The bolometric disk
luminosity is $6 \times 10^{44}$~\ergs, which gives $L$/\Ledd$\approx
0.1$, about a factor of 1.8 smaller than the one assumed in several
previous publications from the campaign. The total {\it observed}
disk$+$X-ray luminosity at HJD=2459322 is about $9.8 \times
10^{44}$~\ergs\, which differs from the bolometric luminosity due to
the disk inclination.  The photoionization calculations described
below assume that this is also the total luminosity seen by the BLR
gas and the disk-wind described below.  The assumed disk+X-ray SED and
the \cosstis\ spectrum are shown in Fig.~\ref{full_SED}.

%

\subsection{Variability}

\subsubsection{Lag measurements}

This work aims to analyze various observed situations involving the
mixing of two sources of lags: X-ray illumination and variable BLR and
DW emission.  The HST/\swift\ data and the derived lags are described
in C23 and L24. Additional ground-based data in the {\it u, g, r, i,}
and {\it z}-bands are taken from Table\,1 and 2 in L24, and from the
analysis of the AGN STORM~2 ground-based photometric campaign by
Montano (in preparation).  We also use our own ICCF lag measurements
using data from C23.

 The C23 results and our measurements were derived from the computed
 interpolated cross-correlation-function (ICCF)
 \citep[e.g.,][]{Peterson2007} with no detrending or smoothing of the
 light curves. The L24 lags are the results of frequency-resolved
 analysis \citep{Uttley2014} and thus depend on the chosen range of
 frequency (given in units of day$^{-1}$).  The differences between
 the two approaches are minor for the lowest frequency bin used by
 L24, which corresponds to 50-100 days. The differences increase at
 higher frequencies that correspond to shorter and shorter periods. In
 particular, the so-called ``Epoch\,1'' and ``Epoch\,2'' lags
 discussed in L24 were derived from a higher frequency bin
 corresponding to 20-70 days. They should not be directly compared
 with the ICCF-derived lags. The shortening lags with increasing
 frequency are similar to the decreasing lags caused by light-curve
 detrending.

 All lags listed in L24 were measured relative to the \swift/UVW2
 band. Since we are after the lags relative to the ionizing continuum
 flux, we re-measured the lag of the UVW2 band relative to the
 1144\AA\ rest-wavelength continuum, which we used as a proxy for the
 light curve of the ionizing radiation. This lag depends slightly on
 the 1144\AA\ flux. Over the first 200 days of the campaign, it is
 roughly 0.6 days. This is identical to the lag of 0.66 days measured
 by C23 within the uncertainties. As shown below, additional lags
 between the 1144\AA\ band and the ionizing flux are also important.

\subsubsection{Changing SED}

C23 conducted a flux-flux analysis to remove non-variable and
long-term variable components from the observed spectra. The method
assumes no wavelength-dependent variations in the intrinsic disk
SED. Given this assumption, they show that the shape of the variable
SED (which includes the disk and diffuse BLR emission) is consistent
with SEDs of optically thick geometrically thin accretion disks.

\begin{figure} \centering

                \includegraphics[width=0.95\linewidth]
                {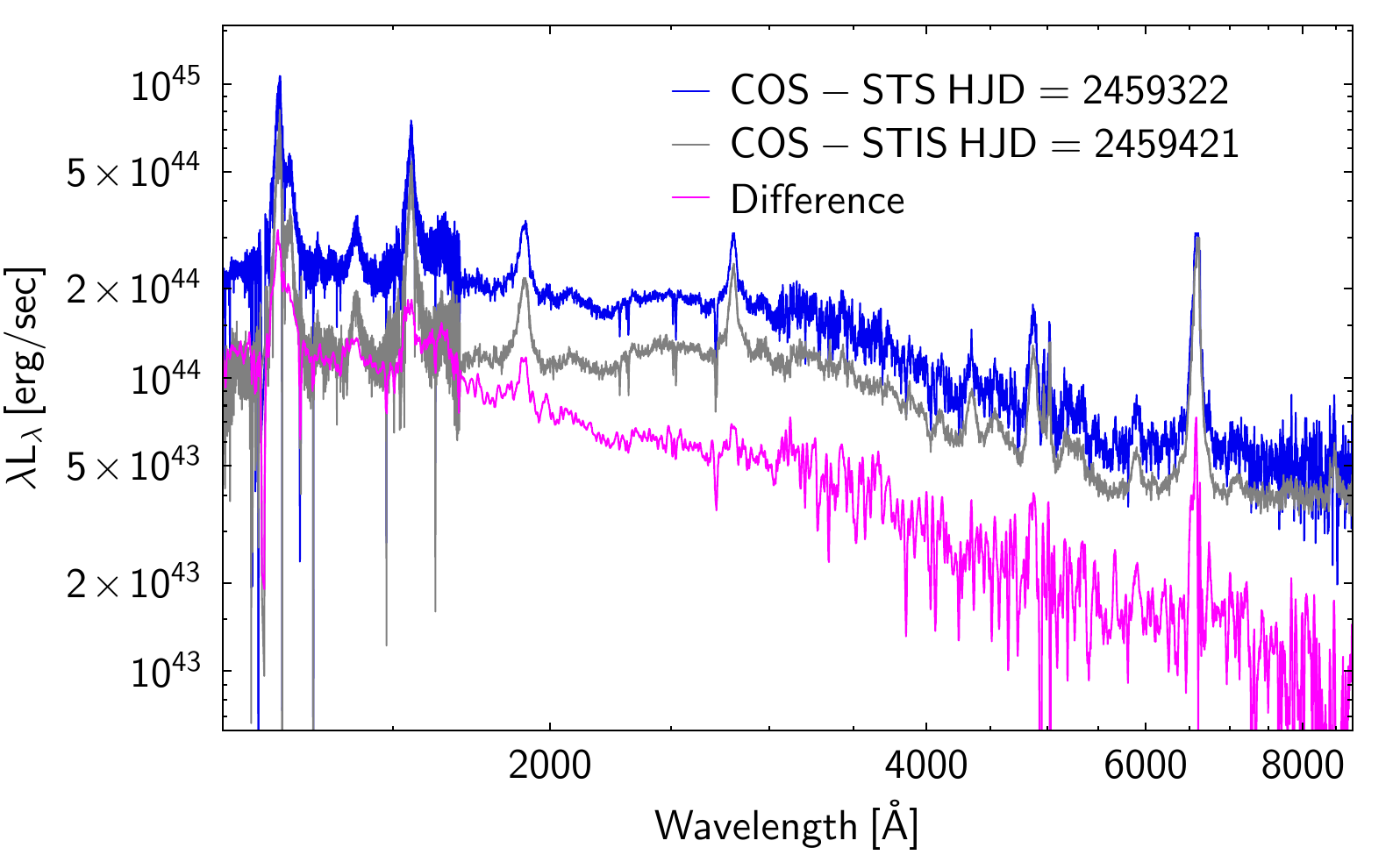}
	\caption{A comparison of two \cosstis\ spectra obtained during
          periods of high (v75, HJD=2459322) and low (v30,
          HJD=2459421) 1144\AA\ flux levels. The magenta curve shows
          the difference between the two spectra.  The similarity of
          the spectra at long wavelengths compared with a factor of
          $\approx 2$ difference in the 1144\AA\ flux suggests a
          significant change in the shape of the SED between the two
          epochs.  }
 
\label{two_cosstis}
\end{figure}

A more detailed examination of the available \cosstis\ spectra
obtained during times of high and low 1144\AA\ flux does not meet this
assumption. This is illustrated in Fig.~\ref{two_cosstis} where we
show \cosstis\ spectra from HJD=2459322 (high flux level) and
HJD=2459421 (very low flux level). Given the negligible contribution
of the host galaxy to these spectra, due to the 0.1 arc-sec slit width
of the HST instruments and the similarity of the two spectra at long
wavelengths, it is evident that the fractional variations at short
wavelengths are larger than those observed at long wavelengths. This
is illustrated in the difference spectrum shown in the figure. The
difference cannot be caused by DC emission from the BLR, which is very
small at short wavelengths around 1144\AA\ and amounts to only 20-30\%
at long wavelengths.

There are several possible ways to explain the significant change in
SED shape illustrated in Fig.~\ref{two_cosstis}. The first is an
additional nuclear source, with a diameter not exceeding 70 pc, that
contributes a significant fraction of the observed flux, especially at
long wavelengths, at times of low AGN luminosity. A compact nuclear
star cluster is one such possibility.  However, we are unaware of a
similar case among local AGN, especially those whose surface
brightness profile was separated into AGN and stellar light components
at HST resolution, as done for this source \citep{Bentz2009}. This
possibility should be further investigated by looking at the expected
spectral characteristics of such clusters, which is beyond the scope
of the present paper. A second possibility is an intrinsic change in
SED shape, which makes the disk bluer when it is brighter. This has
been observed in detailed observations of nearby AGN like NGC\,5548
\citep{Wamsteker1990,Maoz1993,Korista1995}, and in hundreds of low and
high redshift AGN \citep{VandenBerk2004,Sun2014} although other
interpretations have also been proposed \citep{Weaver2022}.

The second possibility mentioned above means that the galaxy light
inside the 5 arc-sec radius \swift\ aperture derived in a flux-flux
analysis overestimates the real flux in the B and V bands by large
factors \citep[for general limitations of the method see][]{Cai2024}.
An estimate of the excess galaxy flux is obtained by comparing the
galaxy flux in the C23 B and V bands with the measured
\cosstis\ fluxes. This suggests an overestimate of $\approx 3.1$ in
the B-band and a factor of $\approx 2.7$ in the V-band. We can also
compare the deconvolved V-band galaxy flux in \cite{Bentz2009} with
the result of the flux-flux analysis. Given the factor of 2.07 in
aperture size between the two observations, this suggests galaxy
V-band flux inside the \swift\ aperture, which is very close to the
one derived from the \cosstis-\swift\ comparison. Given this, the
changing SED shape interpretation is adopted for the rest of this
paper, and we leave a more elaborate analysis for future publications.

Finally, we fitted the fainter \cosstis\ spectrum using a thin
accretion disk SED with a mass accretion rate of 0.05 \Msun/yr, which
is half of our canonical accretion rate. The lower accretion rate disk
reasonably fits the short wavelength SED but underestimates the long
wavelength part. Such a low accretion rate can change the ionization
level non-linearly. All BLR models presented below assume the SED
shown in Fig.~\ref{full_SED}, and a more detailed analysis of this
issue is beyond the scope of the present paper.

\section{Broad line region}
\subsection{RPC BLR models}

Published multi-cloud BLR models that calculate full, multi-line BLR
spectra can be divided into three generic types: constant external
pressure models \citep[see e.g.,][]{Rees1989, Goad1993, Kaspi1999,
  Bottorff2000, Lawther2018}, locally optimally emission cloud (LOC)
models \citep{Baldwin1995, Korista1997,Korista2000, Goad2015,
  Korista2019} and radiation pressure confined cloud (RPC) models
\citep{Baskin2014,Baskin2018, Netzer2020}. The recent dusty-wind
calculations of Czerny, Naddaf, and collaborators, \citep[e.g.][and
  references therein]{Naddaf2023} focus on cloud dynamics and not on
full models of individual sources. Here, we consider only RPC models
and follow the procedures explained in detail in \cite{Netzer2020,
  Netzer2022}.

 Radiation pressure confined clouds are formed where the external
 radiation pressure, due to the incident AGN continuum, exceeds the
 gas pressure in the gas, and the photoionized gas clouds are not
 accelerated outward; i.e., the RPC model is a 1-D (slab) hydrostatic
 solution for the photoionized gas.  No outward acceleration can
 result from either a thick layer of neutral gas at the back of the
 cloud or a situation where the radiation momentum is absorbed by the
 ram pressure of a stationary, low-density medium. RPC clouds are not
 found where the gas pressure is larger than the external radiation
 pressure or where the gas is accelerated outward. We assume that
 there are very few if any, such clouds in the BLR.

The important characteristic of RPC clouds is the distance-independent
ionization parameter at the ionization front inside the
clouds. Because of this, low and high ionization lines are produced
over a very large range of distances in the BLR.  All our
photoionization models were calculated using version 17.02 of the code
$Cloudy$ \citep{Ferland2017} following the exact procedure described
in \cite{Netzer2020}.

The parameters for our RPC model are the central SED, the gas
metallicity, the turbulent velocity inside the clouds, the column
density and covering factor of the clouds, the inner and outer
boundaries of the BLR, $R_{\rm in}$ and $R_{\rm out}$, and the cloud
distribution between the two boundaries. The inner and outer
boundaries determine the lags of the emission lines and the DC
emission relative to the varying ionizing flux. These are the most
important observational constraints on the model.

The overall geometry determines several of the RPC model
parameters. $R_{in}$ depends on the dimensions and physics of the
central accretion disk. For a rotating gravitationally bound cloud
system, the minimal $R_{\rm in}$ is approximately the self-gravity
radius of the disk, which is of order 1000$R_{\rm g}$, where $R_{\rm
  g}$ is the gravitational radius of the BH.  Unbound disk winds can
originate closer to the BH and, while not part of the classical BLR,
can produce observable emission and absorption features. $R_{out}$
depends on the temperature of the nuclear dust, which, following
earlier studies \citep{Netzer2015} and references therein), is assumed
to take the shape of a massive small dusty torus whose inner walls
contain pure graphite grains. Observations of many AGN suggest that
the distance to the inner walls of the dusty torus (the graphite
grains sublimation radius) is about 3--4 times the mean emissivity
radius (MER) of the hydrogen Balmer lines
\citep[e.g.][]{GRAVITY2020b}. For Mrk~817, this corresponds to a
distance of about 100 light days. Given all this, and the observations
described in \cite{Kara2021} and \cite{Homayouni2024}, the overall
range of radii is assumed to be limited to $\log R({\rm
  cm})=15.75-17.5$ ($\sim$ 2 -- 120~light-days). We also assume a
similar inclination of the thin central accretion disk and the thick,
flaring disk BLR. The choice of $\cos i=0.75$ is essential for
calculating the observed versus intrinsic disk flux, the emission line
widths, and the transfer function of the BLR gas.

Below, we focus on three specific BLRs with different distances from
the central BH: Model~1, indicated in all figures by green lines,
represents either the innermost part of the BLR or a ``Disk Wind''
(DW) with $\log R(\rm cm)=15.75-16$ ($\sim 2-4$ light days). The DW
properties are also discussed in
Section~\ref{sec:optically_thick_DW}. The geometry, inclination, and
gas velocity of the DW gas can differ substantially from that of the
flaring-disk BLR.  Model~2, indicated in blue, is an extended
gravitationally bounded BLR with $\log R(\rm cm)=16-17.25$ ($\sim
4-70$ light-days). Model~3 (red lines) is similar to model~2 except
that the entire BLR is shifted to larger radii, $\log R(\rm
cm)=16.25-17.5$ ($\sim 7-120$ light-days).  We also calculated an
additional model (Model~2a) with the same geometry and SED as Model~2
but half the bolometric luminosity. This is used to address the issue
of line and continuum lags during phases of high and low luminosity
and is discussed in Section~\ref{luminosity_dependent_lags}.

All our models assume solar metallicity gas and an internal
microturbulent velocity of 30~\kms. Justification for this choice is
discussed in \cite{Netzer2020, Netzer2022}. The column density of the
clouds in all three models is $N_H=10^{23.5}$\,cm$^{-2}$, which is
large enough to ensure that the total mass of individual clouds is
dominated by neutral gas at the non-illuminated part of the cloud. A
slight deviation from this assumption is discussed in
Section~\ref{sec:specific_DWs}.  All models in this paper assume a
side view of the emitting clouds, i.e., our line of sight to the
central disk is not obscured by clouds.  A factor not included in the
present calculations is the anisotropy of optically thick line
emission, particularly the Balmer lines. For the implications of this
assumption, see \cite{OBrien1994} and \cite{Rosborough2024}.

The covering factor dependence on distance in our BLR models, $C(R)$,
is a crucial parameter. It is given by $dC(R) \propto R^{-\beta} dR$
and the integrated covering factor is fixed by the equivalent width of
the strong UV lines and, as we show later, by the intensity of the
DC. In most models shown here, $C(R_{\rm out})=0.2$. This ``net
covering factor'' can be smaller than the geometrical covering factor
determined by the opening angle of the flaring disk BLR.  We have
experimented with several values of $\beta$ in the range 1.8--2.4,
shown in \cite{Netzer2020} as the range required to explain the
typical observed lags of several strong emission lines. $\beta=2.4$
provides a somewhat better agreement with lag measurements of several
local AGN \citep{Netzer2020}. In all models shown here $\beta=2$.

\subsection{Comparison with the observed spectra}
\label{spectral_fits}

 Our RPC BLR models should be considered representative of many Type-I
 AGN. We ensured that they generally agree with the observed spectra
 of Mrk~817, especially with the measured lags of the strong emission
 lines, and can be used to estimate reliable continuum time delays. We
 did not attempt to calculate accurate velocity-dependent lags of all
 the strong emission lines since such calculations require specific
 treatment of the amplitude and variability pattern of the driving
 continuum light curve. Not all the data required for such modeling is
 available for Mrk~817, and the calculated continuum delay, the center
 of this paper, does not depend much on the emission line model.

\begin{figure} \centering
        \includegraphics[width=0.95\linewidth]{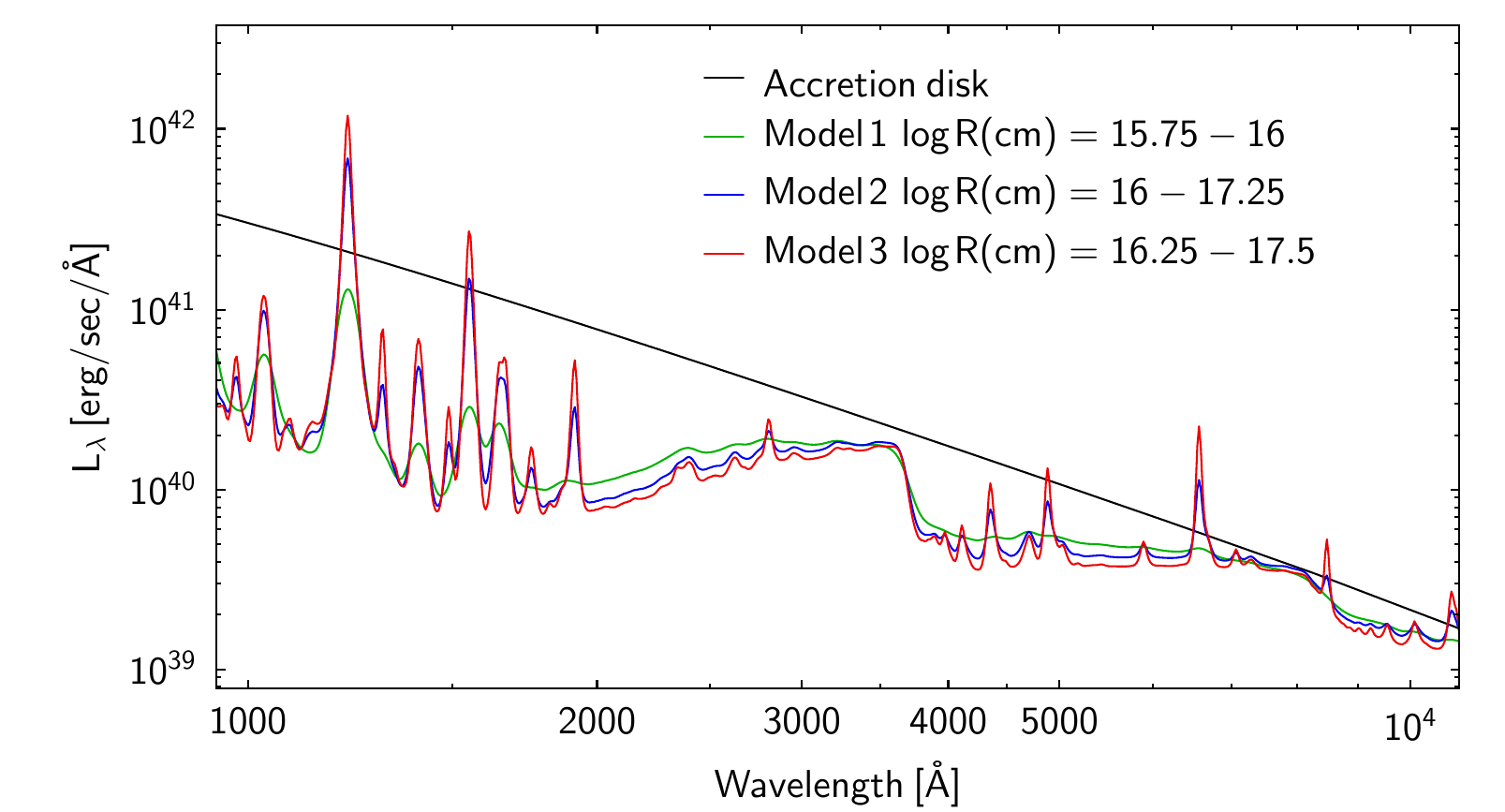}
	\caption{Calculated spectra of three BLR models considered in
          this paper with $C(R_{\rm out})=0.2$. The inner and outer
          boundaries are marked inside the figure. Note the
          significant differences in the emission line widths
          reflecting the different distances from the central BH. In
          particular, in Model~1 (green line), significant velocity
          broadening makes line emission hard to discern.  }
\label{three_BLR_models}
\end{figure}

Fig.~\ref{three_BLR_models} shows the assumed disk continuum and the
calculated spectra of the three generic models. An important feature
is that the fraction of diffuse emission is never smaller than about
10\% of the incident flux, even at the shortest wavelength shown, at
around 1100\AA. As explained below, this is important for fixing the
zero-lag point in broad-band RM observations.

\begin{figure} \centering
        \includegraphics[width=0.95\linewidth]{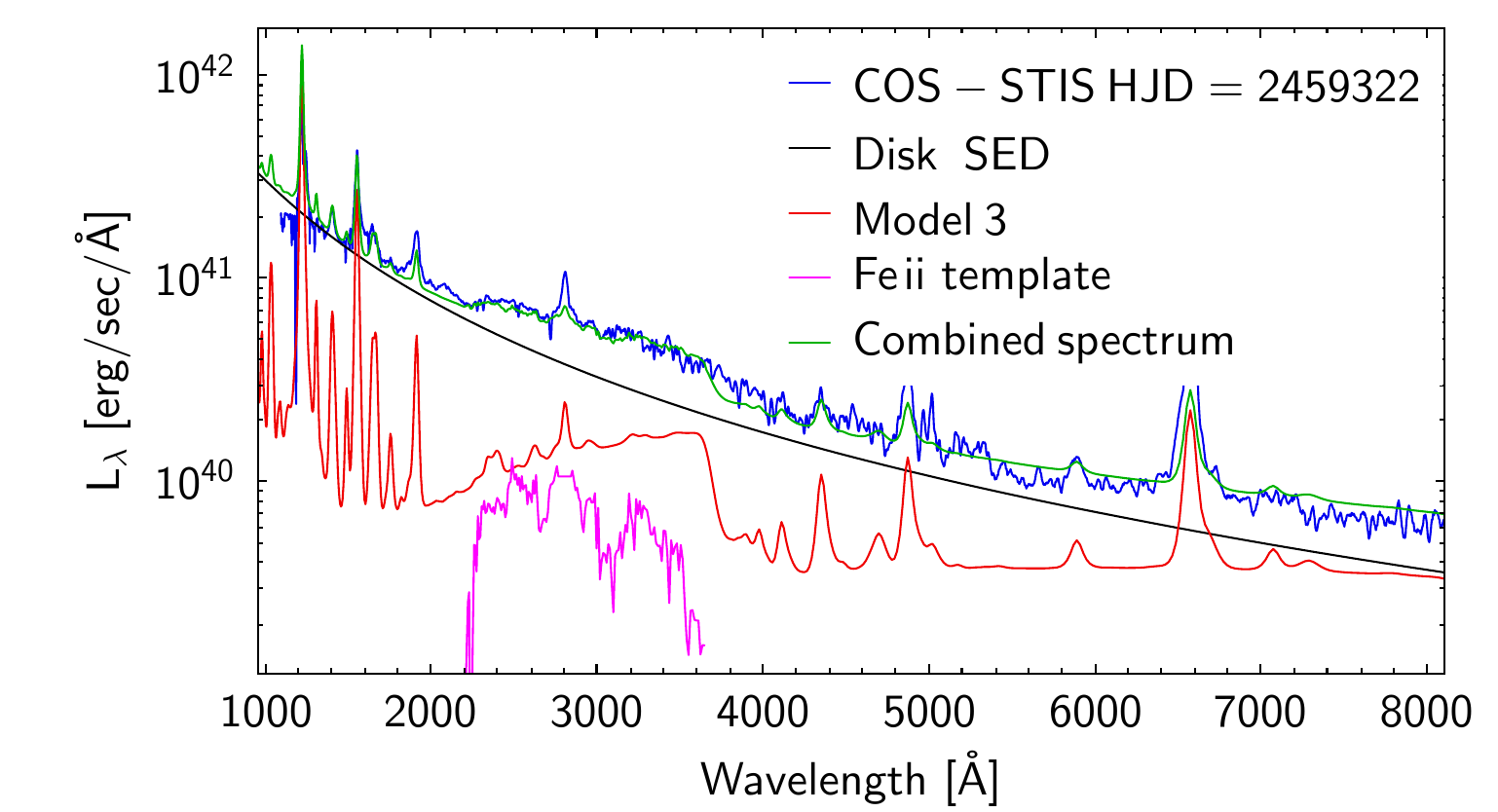}
      \caption{ Comparison of the \cosstis\ spectrum with an empirical
        model (green line) made of two components: Model\,3 from
        Fig.~\ref{three_BLR_models} (red line) and \feii\ template
        based on the UV observations of \cite{Mejia2016} (magenta).  }
\label{cosstis_model3_comparison}
\end{figure}

We compared Model~2 and Model~3 with the \cosstis\ spectrum discussed
in Section~\ref{basic_assumptions}. We focus on Model~3, which we show
later as our preferred model for explaining the observed continuum
lags. The comparison is shown in
Fig.~\ref{cosstis_model3_comparison}. There is good agreement over
much of the 1000--9000\AA\ spectral range, including the strong UV
emission lines and the DC at wavelengths larger than 4000\AA. The
model fails in two important ways: The first is a well-known yet
unsolved problem in modeling the broad Balmer line emission of type-I
AGN \citep[see e.g.,][and references therein]{Netzer2020}. This
extends also to the high-$n$ lines in the Balmer series (see the
deficit flux at around 3800\AA) and was investigated in several
earlier publications \citep[e.g.,][]{Wills1985,Kovacevic2014}.
The second is related to the failure of standard photoionization
models to calculate the intensities of \feii\ lines over the
2000-4000\AA\ range.  This was discussed in numerous publications
\citep[e.g.,][]{Wills1985,Mejia2016,Popovic2019,Ilic2023} where
various empirical recipes were provided to improve the fit of the
data. We artificially added the missing \feii\ flux to
Fig.~\ref{cosstis_model3_comparison} (see figure caption) but not the
high-order Balmer lines, or the optical \feii\ lines. Such additions
will make the overall agreement between model and observations much
better, but they do not mean much given the fundamental failure of the
models.  In the time-lag calculations discussed below, we neglect the
\feii\ lines shown in Fig.~\ref{cosstis_model3_comparison}.

\subsection{Mean emissivity radius (MER)}

\begin{figure} \centering
        \includegraphics[width=0.95\linewidth]{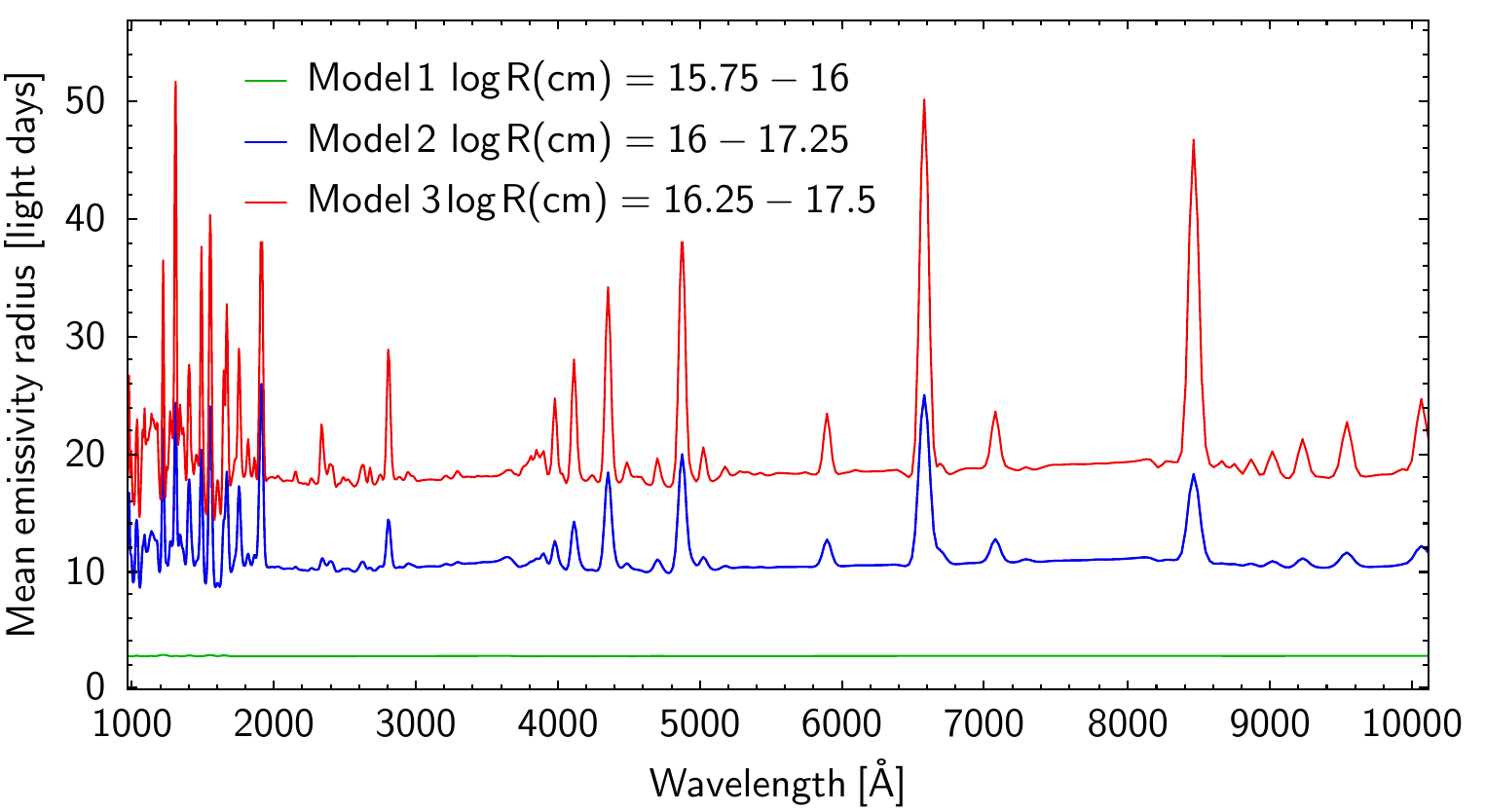} 
	\caption{Mean emissivity radii as a function of wavelength for the three BLR models shown in Fig.~\ref{three_BLR_models}
}
\label{mean_emissivity}
\end{figure}

 Fig.~\ref{mean_emissivity} shows the MERs of the three BLR models
 calculated in this paper. There is a clear difference between the
 MERs of various broad emission lines, which depend mainly on their
 different locations. The MER of the DC, which represents primarily
 bound-free continua with little contribution from free-free emission,
 is almost wavelength-independent.
 We also include Rayleigh scattering from neutral hydrogen, which in the RPC model is seen as very broad and weak \lya\ line wings.  
 
 Several bands used in the observations include individual emission
 lines and DC emissions. In such cases, one can define three somewhat
 different MERs. In principle, the MER of a single isolated emission
 line can be recovered by RM campaigns in cases where the duration of
 the driving continuum pulse is at least as long as the crossing time
 of the line-emitting region (in reality, this is hard to achieve as
 explained in the numerical simulations part in section~4). This MER
 refers to the total line flux. Velocity-resolved lags, when
 available, provide MERs for every velocity component, provided the
 line emission can be separated from the underlying DC emission. The
 third type of MER is important when using broad-band observations
 that combine line and DC emission. Time lags based on such
 observations differ from the other two types of MER. For example, a
 broad-band measurement of \hb\ together with the 4861\AA\ Paschen
 continuum emission will result in a lag that is shorter than the
 measured lag of \hb\ since the MER of the DC is about half the MER of
 the strong Balmer lines. This situation is illustrated in
 Fig.~\ref{MER_near_hb} in the appendix.  As discussed in the
 simulation section below, RM campaigns hardly ever result in lags
 that are longer than about 80\% of the crossing time of the MER. For
 the 1144\AA\ driving continuum used here, the number is closer to
 70\%. These numbers are listed as footnotes in Table~\ref{table_1}.

\begin{table}
	\centering
	\caption{MERs in light days and predicted and observed lags in days.}
	\label{tab:calculated lags}
	\begin{tabular}{lccr} 
		\hline
    Model & 1$^a$ & 2$^b$  & 3$^b$ \\
		\hline
    \hb\ MER & 2.8 & 24.1 & 38 \\
    \hb\ predicted lag & 2.5 & 15 &25\\
    \hb\ observed lag$^c$ & 24  & 24  & 24 \\
    \hline
    \civ\ MER & 2.8 & 18.8 & 31 \\
    \civ\ predicted lag &2.5 & 12.2 &20.1\\
    \civ\ observed lag$^d$ & 12 & 12  & 12 \\
    \hline
    \heii\ MER & 2.8 & 11.0 & 20.7 \\
    \heii\ predicted lag &2.5 & 7.2 &13.5\\
    \heii\ observed lag$^d$ & 9 & 9  & 9 \\
    \hline
 DC MER &2.8  &11.2  &19 \\
 DC predicted lag & 2.5 & 7.3 &  10.4 \\ 
 \hline
 \label{table_1}
	
 $^a$ Predicted lag=MER$\times 0.9$\\
 $^b$Predicted lag=MER$\times 0.65$\\
 $^c$ \cite{Kara2021}\\
 $^d$ \cite{Homayouni2023}
 \end{tabular}
 \end{table}

 The calculated lags for the total \hb\ and \civ\ emission lines in
 Mrk~817, in models~2 (24 and 16 days, respectively), agree with the
 results of \cite{Kara2021}.  Note also that the outer radius in this
 model is about three times the MER of the \hb\ line. All predicted
 line lags in model~3 are also in reasonable agreement with the
 observations. These lags are roughly consistent with canonical lags
 observed in low Eddington ratio AGN \citep[e.g.,][]{Du2019,
   Lira2018}. However, model~1 predicts much shorter lags that agree
 with some \civ\ lag measurements but not with others
 \citep[e.g.,][]{Homayouni2024}. As shown below, this is the main
 reason for the association of this model with a disk wind.

\subsection{Combined disk-BLR models}

The basic assumption in this paper is that the measured broad-band
UV/optical lags are combinations of short lags of order 0.5-2 days,
from an X-ray illuminated disk, with longer lags typical of BLR
emission.  A rigorous way to combine the two is to calculate transfer
functions (TFs), or response functions, that combine the BLR and disk
emission. This is tricky since every line in the BLR has a different
TF, and some line emission is mixed with a DC emission with additional
TFs. Earlier papers have tried this idea using a generic TF
representing a ``typical'' BLR response. One such case is presented in
L24. As shown in the appendix, this assumed TF disagrees with the DC
TF computed here.

Here, we adopt a time-domain forward modeling approach that was tested
successfully in earlier studies of other AGN.  The combined disk-BLR
lag, which does not include delays due to thermal dust emission (to be
treated later), is a wavelength-dependent expression given by:
\begin{equation}
 \tau_{\lambda,{\rm tot}} =
 \tau_{\lambda,{\rm irr}}  \left[ \frac{L_{\rm inc}}{L_{\rm tot}} \right]
  + 
 f_{\rm lag}({\rm TF}) \times \tau_{\lambda,{\rm diff}} \left[ \frac {L_{\rm diff}}{L_{\rm tot}} \right] \,\, {\rm days} 
 \label{tau_tot}
\end{equation}
where $\tau_{\lambda,{\rm irr}}$ and $\tau_{\lambda,{\rm diff}}$ are
the lags due to the illuminated disk and the diffuse BLR emission,
respectively, $L_{inc}$ is the luminosity of the incident continuum,
$L_{\rm diff}$ is the diffuse (BLR) luminosity that includes DC and
broad emission lines, $L_{\rm tot}=L_{\rm diff}+L_{\rm inc}$ and
$f_{\rm lag}({\rm TF)}$ is a correction factor that depends on the
transfer function (TF) and nature of the driving light curve.  The
above expression, with $f_{\rm lag}({\rm TF})=1$, is taken from
\cite{Netzer2022} and similar earlier expressions are discussed in
e.g., \cite{Lawther2018} and \cite{Hernandez2020}.  The numerical
simulations below show that the best value for Models\,2, and 3 is
$f_{\rm lag}({\rm TF})$=0.5.

  The total time lag calculated in this way, $\tau_{\lambda,{\rm
      tot}}$, should be compared either with the lags measured in the
  lowest frequency bins in the frequency-resolved analysis of
  \cite{Cackett2022} and L24 or the lags measured with the ICCF method
  for non-detrended LCs. The disk-lag term, $\tau_{\lambda,{\rm
      irr}}$, is given by Eqn.1 of \cite{Netzer2022} and is basically
  identical to Eqn.~12 in \cite{Fausnaugh2016}. It assumes $X=2.49$,
  L$_x$/\Ldisk=0.1, $h=10 R_{\rm g}$, disk albedo of 0.2, and no
  relativistic corrections; all standard assumptions used in earlier
  calculations.


%

\begin{figure} \centering
        \includegraphics[width=0.95\linewidth]{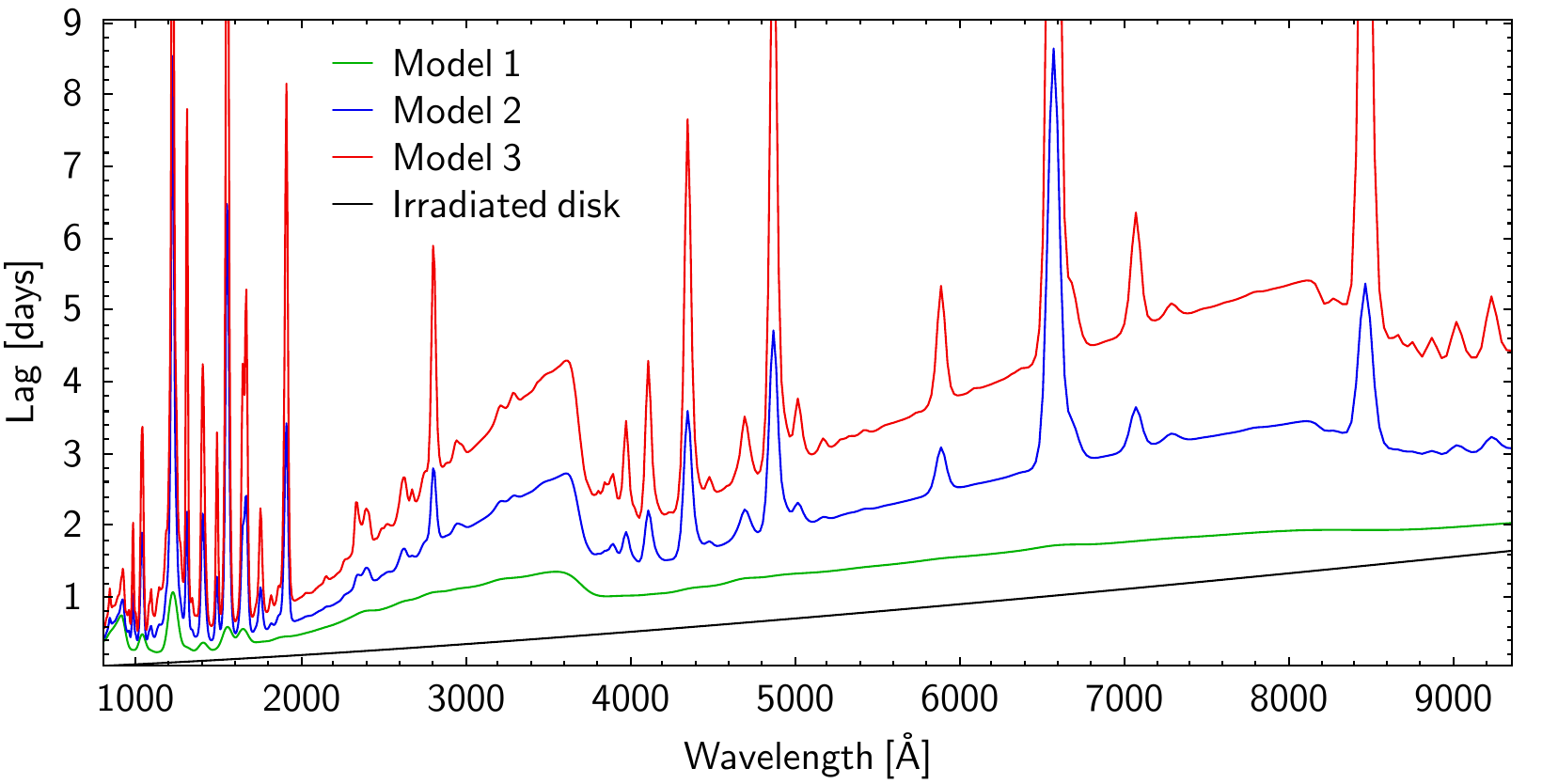}
	\caption{Computed continuum lags for the three models shown in
          Fig.\ref{three_BLR_models} assuming $f_{lag}=0.9$ for
          model~1 and $f_{lag}=0.5$ for the other two models (see text
          for justification). The large differences in lags mostly
          reflect the different MERs of the models.  }
\label{continuum_lags_1}
\end{figure}

Fig.~\ref{continuum_lags_1} shows computed lags for the three cases
shown in Fig.~\ref{three_BLR_models}. The numerical simulations below
show that the best estimates require $f_{\rm lag}=0.9$ for Model\,1
and $f_{\rm lag}=0.5$ for Models\,2, and 3.  The total disk+BLR lags
are very different at all wavelengths longer than about 1800\AA\,
reflecting the different MERs of the three models.

\section{Simulations}
\label{sec:simulations}

\subsection{The dependence of the measured delay on DC fractional contribution}

Before proceeding to use eqn.~1, we want to estimate the effect on the
measured delay of the differing diffuse emission contributions
(assumed here to be only DC emission) to the total light,
DC/(disk$+$DC), for our model BLR. The first step is to investigate
the behavior of a simple toy model in which effects due to geometric
dilution of the input signal (resulting from a spatially extended BLR)
are completely removed. This we achieve by modeling the response of a
geometrically thin ring of gas located 10~light-days from the
continuum source and viewed face-on. For this geometry, the response
function may be represented by a delta function in time. Assuming a
linear response approximation, the output signal from this thin ring
of gas (the DC component) will be a copy of the input signal (the
driving component, initially taken to be the 1144\AA\ continuum
light-curve for the full campaign), shifted in time by an amount equal
to the light-crossing time to the region. The measured delay, as
determined by cross-correlating the input signal with the output
signal, will, in this instance, be precisely 10 days.

Next, we investigate the measured delay between the input signal and
an output signal comprising the sum of the input signal and the
responding DC component, such that the fractional contribution of the
DC component to the total light is in the range 0--1.  The delay is
computed from the ICCF centroid, here determined for time intervals
spanning values of the cross-correlation coefficient $> 0.8\times {\rm
  CCF}_{\rm peak}$, where CCF$_{\rm peak}$ is the peak correlation
coefficient.

\begin{figure}
    \centering
    \includegraphics[width=\columnwidth]{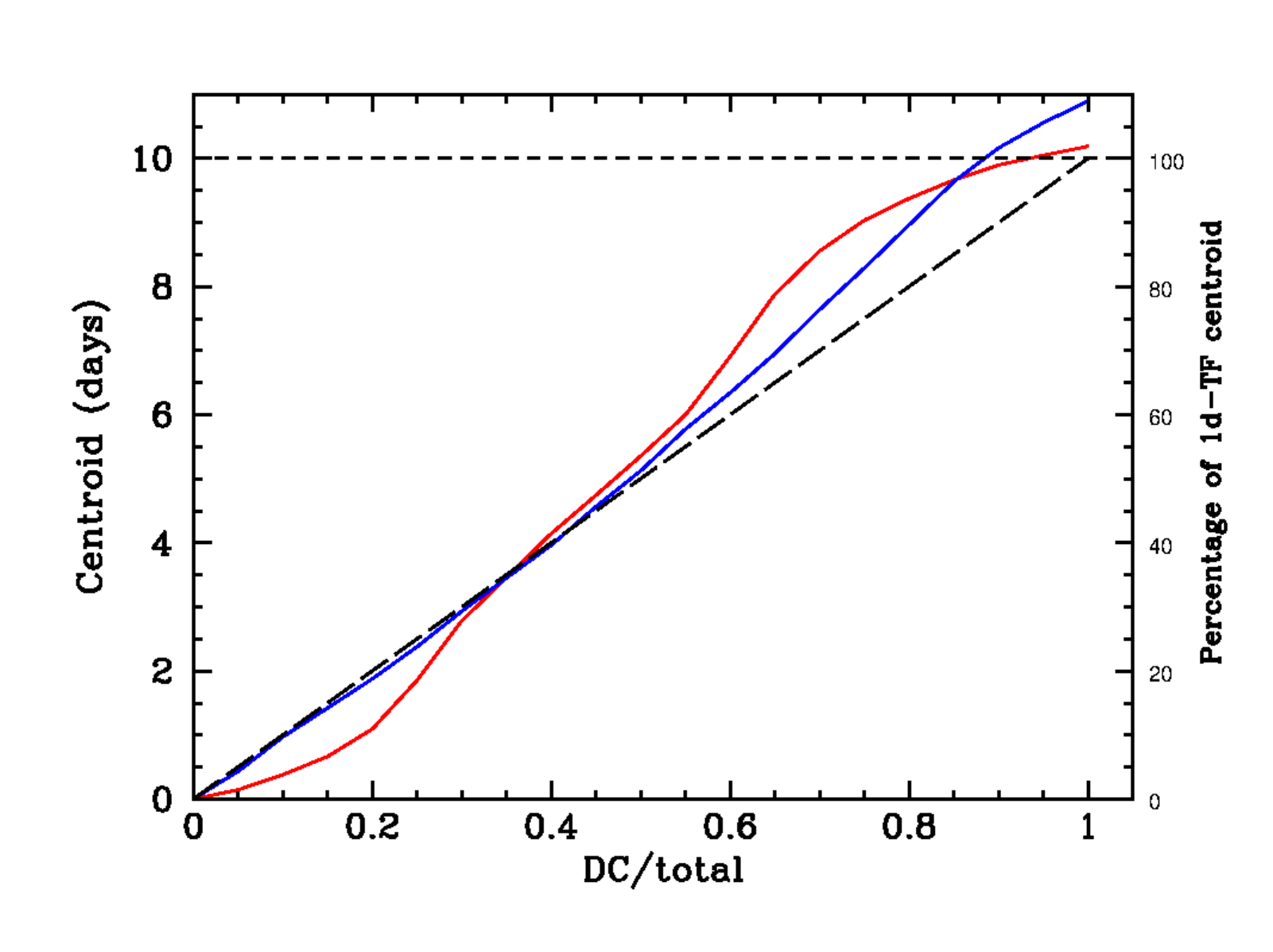}
    \caption{The measured delay (ICCF centroid) as a function of DC
      fractional contribution to the total light (Disk$+$DC), for a
      face-on thin ring with a radius of 10 light days. The output
      signal is the sum of the input signal (driving light curve) and
      responding signal, such that the responding signal contributes a
      fraction of the total light in the range 0--1. The measured
      delay is taken to be the centroid of the cross-correlation
      function determined from cross-correlating the driving and
      responding signals (see text for details). The dashed horizontal
      line indicates the delay of the DC component only
      ($=10$~days). The diagonal line indicates linearity -- the
      measured delay is directly proportional to the fractional
      contribution of the DC component to the total light. Results are
      shown for two instances of the driving continuum light-curve:
      F(1144\AA) light-curve for Mrk~817 (red), and a damped random
      walk, $T_{\rm dur}=500$~days and damping timescale $T_{\rm char}
      =40$~days (blue).}
    \label{thin_ring}
\end{figure}

The results from this study are shown in Fig.~\ref{thin_ring}, solid
red line. As indicated in Fig.~\ref{thin_ring}, the measured delay,
with respect to the input signal, increases as the fractional
contribution of the DC component to the total light
increases. However, the relationship, while approximately linear, is
not strictly linear, even for this simple model, with the measured
delay being biased toward the delay of the major contributory
component. Since the measured delay also depends on the characteristic
variability timescale $T_{char}$ of the driving continuum relative to
the size of the region being probed, and light-curve duration (see
Appendix~D), a strictly linear dependence between measured delay and
DC fractional contribution, may in practice never be realized. We have
tested this behavior using simulated driving light curves of fixed
duration but with different characteristic damping timescales. The
measured lag dependence on DC fractional contribution for a face-on
thin ring is close to linear at small DC fractional contributions if
$T_{\rm char}$ is larger than the MER, provided that the light-curve
is of sufficient duration. Fig.~\ref{thin_ring} shows an example
simulation using a driving light-curve with $T_{\rm dur}=500$~days and
damping timescale $T_{\rm char}=40$~days (see Appendix~D for details).

\begin{figure}
    \centering
    \includegraphics[width=\columnwidth]{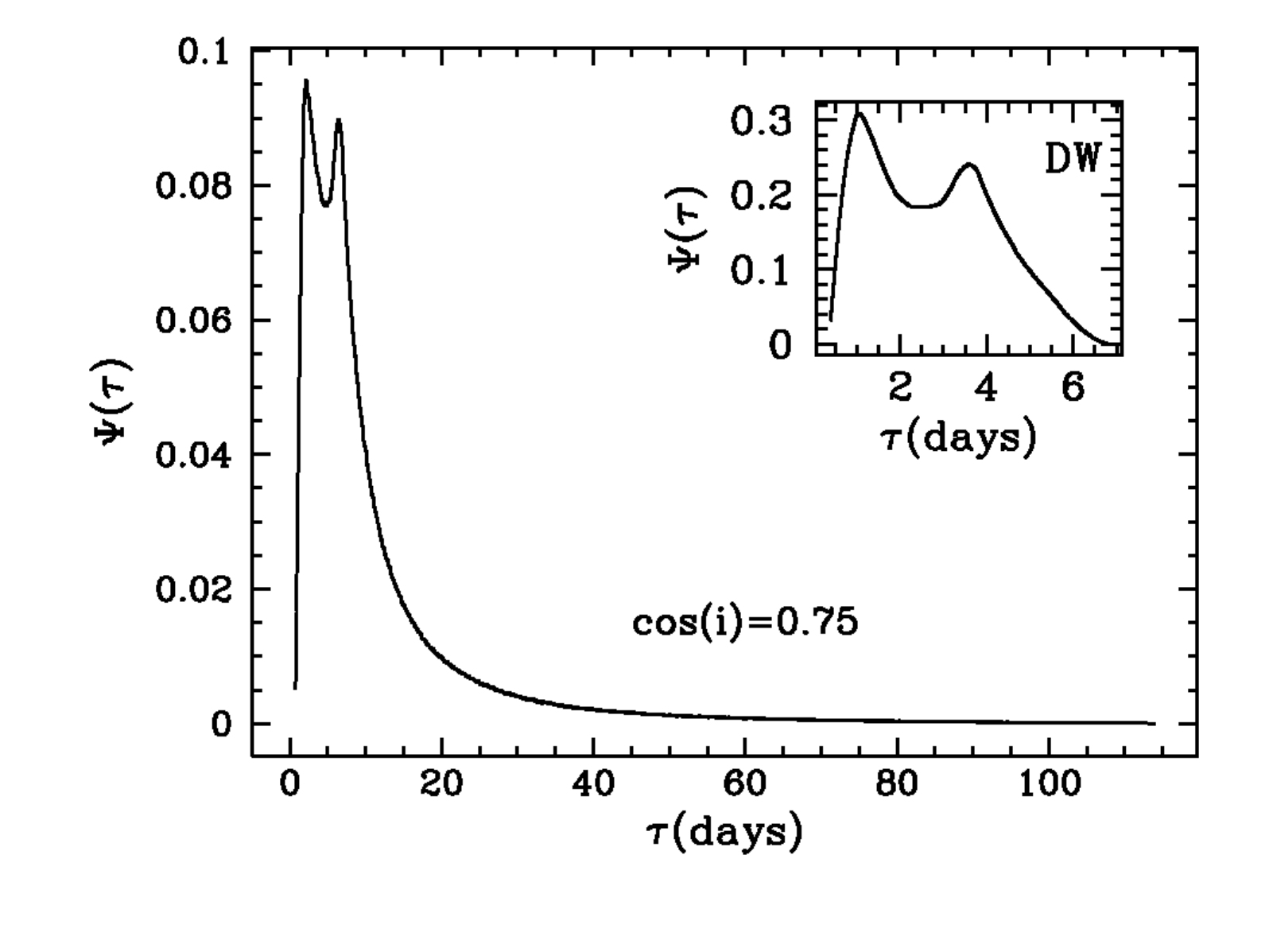}
    \caption{The 1-d response function (or TF) for the
      7995\AA\ continuum band, for our model BLR viewed at line of
      sight inclination $\cos(i)=0.75$. The response function has been
      normalized to unit area. Inset -- the 1-d response function for
      7995\AA\ continuum band arising from the DW (Model\,1) viewed at
      the same inclination.}
    \label{transfer_fn}
\end{figure}

\begin{figure}
    \centering
    \includegraphics[width=\columnwidth]{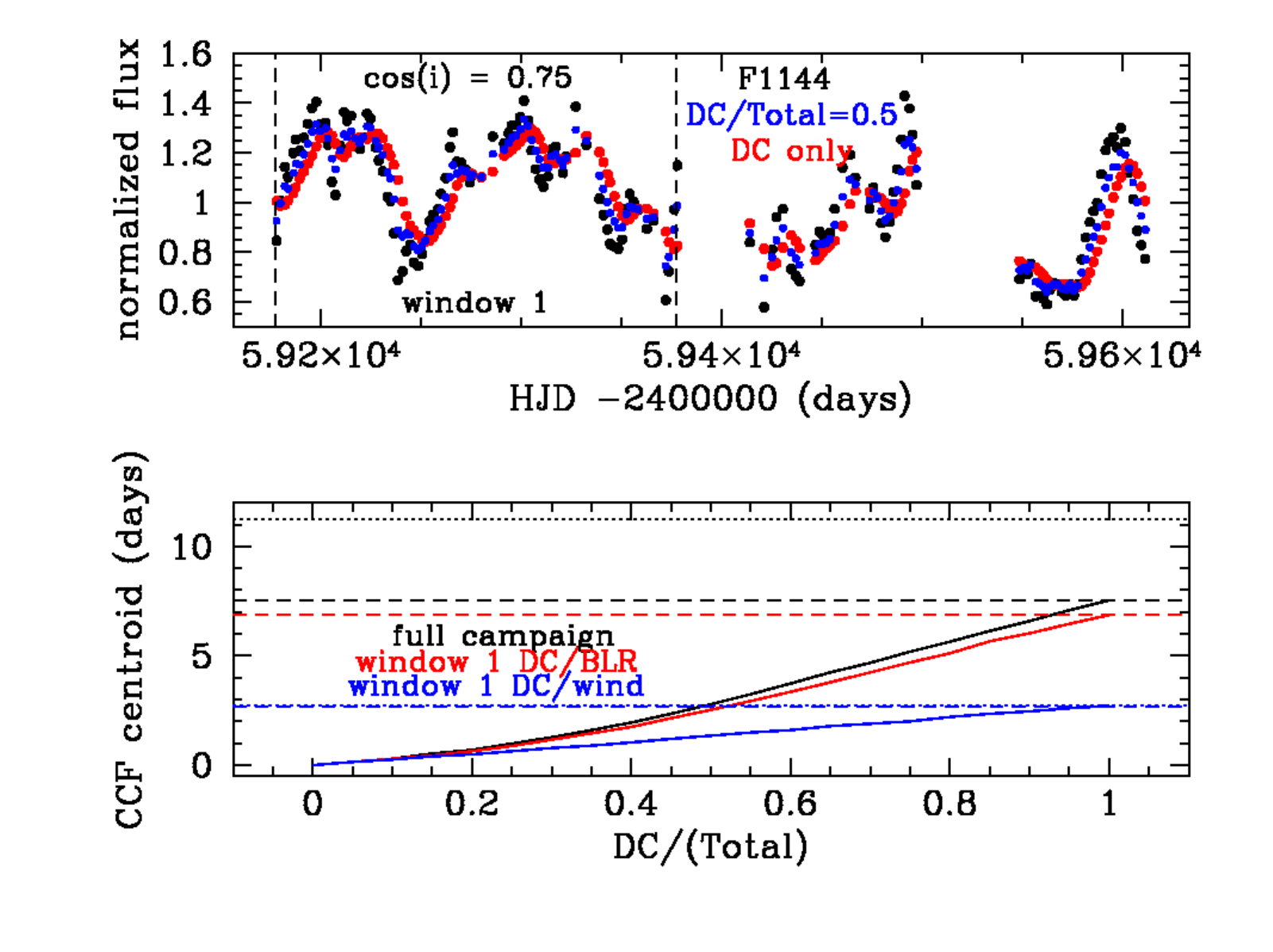}
    \caption{Upper panel -- input driving continuum (F1180) for
      Mrk~817 (black), and output light-curves with fractional DC
      contributions DC/(disk+DC) = 0.5 (blue), and DC only (red). Each
      light curve has been normalized to its mean value to aid
      clarity. The location of "window 1" (HJD=2459177-2459377) is
      indicated by the vertical dashed lines.  Lower panel -- The
      measured delay (ICCF centroid) as a function of DC fractional
      contribution to the total light for the full light curve (black)
      and "window 1" (red). Also shown is the measured delay for the
      wind component against its fractional contribution to the total
      light (blue) without any DC contribution from the BLR.  The
      horizontal dashed lines indicate the measured delay of the DC
      component only for the full campaign (black, $\approx
      7.5$~days), window 1 (red, $\approx 6.9$~days) and window 1 for
      the wind component only (blue, $\approx 2.7$~days). The
      horizontal dotted lines indicate the MER for Models\,2 and 3
      (black, 11.3~days), and for Model\,1 (blue, 2.75~days).  }
    \label{lcurve_lag}
\end{figure}


\subsection{Mrk~817}

For Mrk~817, we model the response of the diffuse continuum band at
7995\AA\, arising from a BLR geometry here described as a
geometrically-thick flared disk, with inner radius $R_{\rm
  in}=10^{16.0}$~cm, outer radius $R_{\rm out}=10^{17.25}$~cm,
semi-opening angle 11.4~degrees, and differential radial covering
fraction $dC(R) \propto R^{-2}dR$. The total covering fraction for
this particular geometry, $C(R_{\rm out})$, is 0.2. The response
function for the 7995\AA\ DC band for our adopted geometry viewed at a
line-of-sight inclination $\cos(i)=0.75$ is shown in
Fig.~\ref{transfer_fn}. The response function is double-peaked,
characteristic of a flattened geometry, and is strongly peaked toward
smaller delays, with a weak extended tail toward longer delays. The
MER or, equivalently, the centroid of the response function for this
continuum band is 11.26~days.

For the proxy driving continuum, we use the 1144\AA\ continuum band
located short-ward of Ly-$\alpha$. We choose this band due to its
proximity to the EUV continuum responsible for driving continuum and
emission-line variations at longer wavelengths, and because at shorter
wavelengths, far enough from the Rayleigh scattering feature beneath
Ly-$\alpha$, the DC contribution to the total light is small when
compared to DC contributions at longer wavelengths.

We use the part of the 1144\AA\ continuum LC extending from
HJD=2459177 to HJD=2459377 (hereafter ''window 1''). This covers the
period from the beginning of ''window~1'' to the end of ''window~3" as
defined in \cite{Homayouni2024}. We expect some dependence of the
results from the simulations on our chosen window.  \textbf{} As for
our simple toy model, we drive the response of the diffuse continuum
band at 7995\AA\ with our proxy driving continuum and measure the
delay between the driving continuum and diffuse continuum band from
the ICCF. The measured delay for the DC component only ($=7.28$~days)
over the full campaign is $\approx$35\% smaller than the model
response function centroid for this band ($=11.26$~days). While the
ICCF centroid is commonly taken to be representative of the
emissivity-weighted radius (or equivalently response function
centroid), previous studies have shown that this is rarely realized in
practice (e.g., Goad and Korista 2014; 2015), a combination of a
mismatch between the characteristic timescale of the driving continuum
and BLR "size", short campaign duration (see Appendix D), as well as
the weak extended tail of the response function.

We next investigate the delay between the proxy driver and the sum of
driver $+$ diffuse continuum band light curve, where the fractional
contribution of the diffuse continuum band light curve to the total
light varies from 0--1.  Example light-curves for the driving
continuum (F(1144\AA), black points), and driver$+$DC with DC
fractional contributions to the total light, DC/total$ = 0.5$ (blue),
and DC only (red), normalized to their mean values, for the full
campaign are illustrated in the upper panel of
Fig.~\ref{lcurve_lag}. Note that as the DC contribution increases, the
variability timescale increases and the variability amplitude
decreases as expected. In the lower panel of Fig.~\ref{lcurve_lag} we
illustrate the delay dependence on DC fractional contribution to the
total light for both the full campaign (black) and window 1
(red). Once again, the measured delay increases as the DC contribution
to the total light increases.

We have also investigated the same scenario for our Model~1 (small BLR
or a DW), where the inner and outer boundaries are much closer to the
BH, $\log R({\rm cm})=15.75-16$.  The response function for this case
is shown in Fig.~\ref{transfer_fn} and as a blue line in
Fig.~\ref{lcurve_lag} . The situation here is more similar to the thin
ring case. The measured delay increases with increasing fractional
contribution in an approximately linear manner. The mean reduction in
lag relative to the measured MER ($\approx 2.6$~days) is only about
10\%, a consequence of the small MER relative to $T_{\rm char}$.

\subsection{General recommendations}

To summarize, numerical simulations applied to the present Mrk~817
campaign suggest that the approximation given by equation 1 with
$f_{\rm lag}=1$, overestimates the lag induced by the variable BLR by
a geometry-dependent factor. The reasons are the exact nature of the
driving continuum, the uncertainties on the TF, and the known
discrepancy between the MER of the BLR and the duration and
characteristic time scale of the variable ionizing continuum. A
first-order correction for Models~2 and 3 would be multiplying
$\tau_{\lambda,{\rm dif}}$ in equation~1 by a factor $f_{\rm
  lag}\approx 0.5$. For Model~1 with the much smaller BLR, we find a
factor $f_{\rm lag}$ of about 0.9. As for other AGN, modeling a
specific BLR using measured line and DC lags could enable better
calculations of the TF and lead to improved estimates of this
factor. In the following sections, we use $f_{\rm lag}=0.5$ for
Models\,2\ and 3 and $f_{\rm lag}=0.9$ for Model~1.

\section{Results and discussion}

\subsection{Long-term BLR size and continuum lags}

Having discussed the uncertainties on the two terms in
eqn.~\ref{tau_tot}, we can now compare Models\,2 and 3 with the
measured HST/\swift\ time lags.  At this stage, we focus on mean
observed properties and do not consider luminosity variations
discussed in section~\ref{luminosity_dependent_lags}.  The data we
compare are the lags computed by C23 using the ICCF method (C23
Table~2) and those derived by L24 using the frequency-resolved method
(L24 Table~1). The latter were obtained over a frequency range of
0.01--0.02 day$^{-1}$. The two sets of lags are identical within the
errors, and we used their average values in the following discussion.

As explained earlier, we chose the 1144\AA\ rest-wavelength continuum
as our shortest wavelength point and added 0.6 days to all lags
calculated relative to the \swift/UVW2 band. In addition, we
considered the expected lag between the 1144\AA\ band and the ionizing
continuum. This is illustrated in Fig.~\ref{three_BLR_models} that
shows that the combination of broad line wings, weak emission lines,
the Rayleigh wings of the \lya\ line, and residual Balmer continuum
emission result in total diffuse emission of roughly 10\% of the
incident continuum radiation.  This translates to an additional lag
relative to the ionizing continuum of about 0.5 days. Since we do not
have direct observations of the ionizing continuum, we added this
theoretically-derived lag to the measured lags in C23 and L24.

\begin{figure} \centering
        \includegraphics[width=0.95\linewidth] {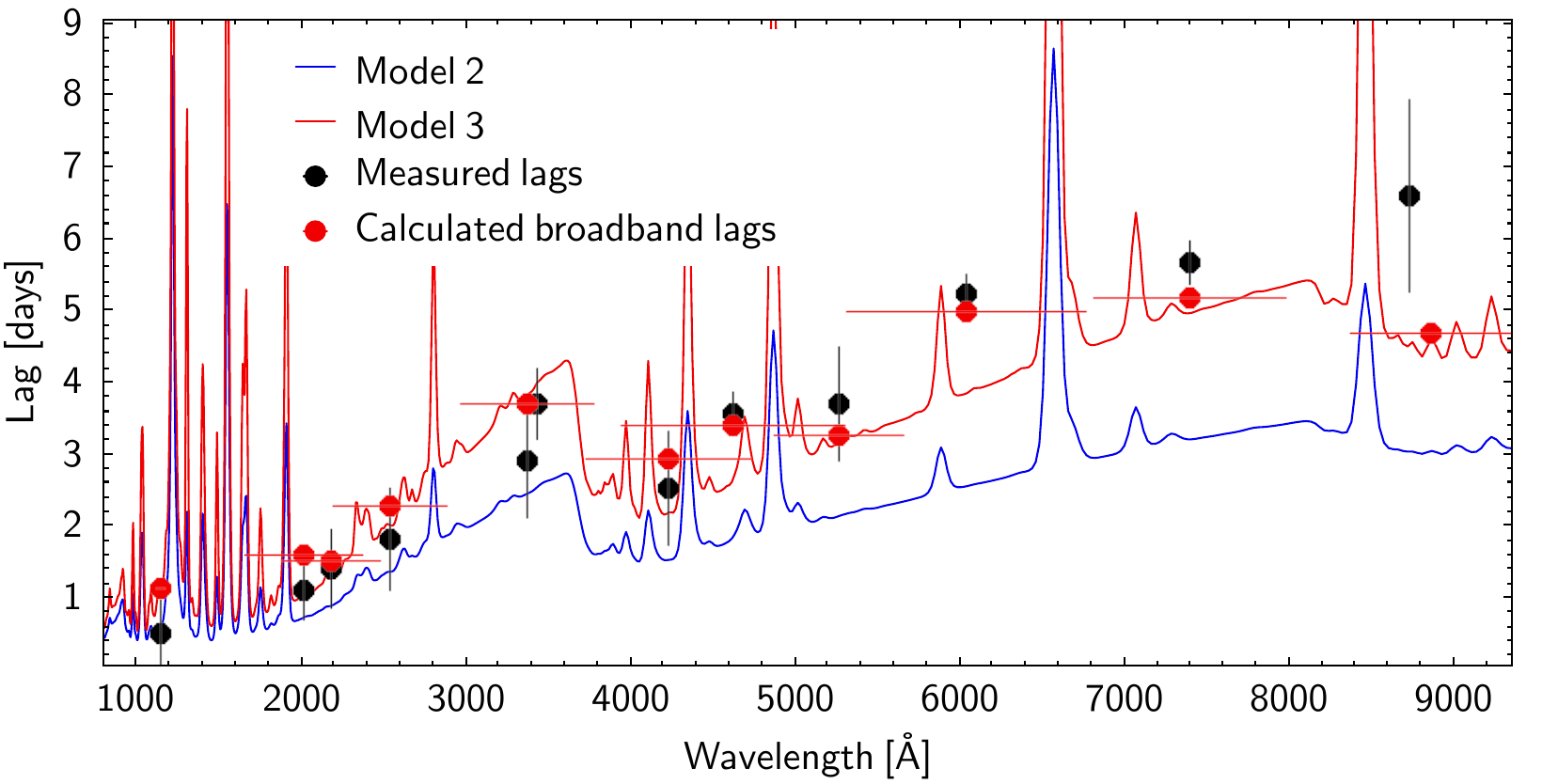}          
	\caption{Computed and observed lags for the entire
          campaign. Data are from C23 and L24, assuming a lag of 0.6
          days between the HST/1144\AA\ and \swift/UVW2 bands. An
          additional lag of 0.5 days is added to all bands to allow
          for the lag between the ionizing continuum and the
          HST/1144\AA\ band (see text for details). The horizontal red
          lines show $1/e$ wavelength limits on the \swift\ and
          ground-based bands. The BLR models are identical to the ones
          shown in similar colors in Figs.~\ref{three_BLR_models},
          \ref{mean_emissivity}, and \ref{continuum_lags_1}.  }
\label{observed_and_calculated_lags_model2_model3}
\end{figure}

Fig.~\ref{observed_and_calculated_lags_model2_model3} shows a
comparison between the model and observations.  The best agreement is
with model~3 with a BLR that extends from $\log R({\rm cm})=16.25$ to
$\log R({\rm cm})=17.5$ (6.9-122 light-days).  The biggest difference
between the model and observations is the $z$-band lag. As we show in
the disk-wind sections below, at least part of it can be explained by
even longer lags due to the dust in the torus \citep[see][figure
  2]{Netzer2022}.  The predicted and observed line lags listed in
Table~1 prefer Model~2 over Model~3, but final measurements of the
\hb\ lags are not yet available, and the \civ\ lags are changing
dramatically over time \citep{Homayouni2023}. Given all the above,
Model~3 is our preferred BLR model.

\subsection{Luminosity dependent lags}
\label{luminosity_dependent_lags}
Mrk~817 is characterized by large luminosity variations over
relatively short periods.  This must lead to model-dependent lag
variations during the campaign.  The measured lags during high-and-low
periods depend on luminosity, the disk SED, and the nature of the
driving light curve. Detailed modeling requires information about the
time-dependent ionizing SED, which is not available except for partial
information through the analysis of the X-ray observations.  The
observed change of color (''bluer when brighter'') possibly indicates
changing accretion rates, but the time scale is very short compared to
the viscosity time of thin accretion disks. Given this complexity, we
experimented only with a simple scenario where $L_{\rm bol}$ changes
by a factor of two (the largest variation observed in the
1144\AA\ continuum light curve) without a change in the shape of the
SED.

Model~2a represents the same BLR assumed earlier exposed to an
incident continuum with the same SED as in Model~2 but with $L_{\rm
  bol}$, which is a factor of two smaller. Assuming $R_{\rm in}$ does
not change, and an outer boundary which is determined by the
sublimation of graphite grains, $R_{\rm out} \propto L_{\rm
  bol}^{1/2}$. Thus $R_{\rm out}({\rm Model~2})= \sqrt 2 \times R_{\rm
  out}({\rm Model~2a})$. We also include a small variation (2.5\%) to
the covering factor, which is slightly smaller in Model\,2a because of
the smaller $R_{\rm out}$.

Table~\ref{table_2} compares line and continuum lags in Models~2 and
2a. None of the emission line lags follow the predicted $\sqrt 2$
reduction in time lag. Lines more efficiently emitted in the outer
BLR, like \hb, are closer to this prediction. Other lines, like
\heii$\lambda 1640$, are coming from the innermost BLR, and their MER
hardly changes despite the large change in line luminosity. This is
also the case with the DC emission. These tendencies are even clearer
when examining lag variations in different parts of the line profiles
(not shown here). The lags of the line cores emitted by the outer BLR
change the most while the far wings, 4000\,km\,s$^{-1}$ from the line
center, are emitted much closer to the ionizing source, where the
structure of the line emitting zone hardly changes.

\begin{table}
	\centering
	\caption{Luminosity dependent MER (light days).}
	\label{tab:calculated lags_changes}
	\begin{tabular}{lccr} 
		\hline
    Model &  2 $^e$   & 2a $^f$ & Ratio  \\
		\hline
    \hb\ core      & 29.0  & 21.6  & 1.34\\
    \hb\ 4000~\kms\ & 13.7  & 10.9  & 1.26 \\
    \hb\ total     & 24.1  & 18.6  & 1.30 \\
    \hline
    \civ\ core      & 24.6 & 19.0  & 1.29 \\
    \civ\ 4000~\kms\ &10.5  & 10.0  &1.05  \\
    \civ\ total     & 18.8 & 15.1  & 1.26 \\
    \hline
    \heii\ total    & 11.0 & 10.1  & 1.09 \\
    \hline
    DC              &11.2  &10.0   &1.12   \\
 \hline
 \label{table_2}	
 $^e\log R({\rm cm})=16-17.25$\\
  $^f\log R({\rm cm})=16-17.10$
 \end{tabular}
 \end{table}

We have also examined a similar situation for Model~3 by defining
Model~3a where $L_{\rm bol}$ is reduced by a factor two, and the outer
radius changed from $\log R({\rm cm})=17.5$ to $\log R({\rm
  cm})=17.35$. The relative decreases in the various MERs are similar
to those found when comparing Model~2 to Model~2a (see
Table~\ref{table_2}).

  Finally, we have experimented with a model with smaller
  $R_{in}$. Such a case results in larger changes in line and DC
  lags. However, there is no simple mechanism to justify such an
  assumption, except perhaps the possibility that small optical depth
  clouds may become optically thick during lower luminosity phases.


%

\subsection{Disk winds (DWs) and BLR obscuration}
\label{sec:DW}

\cite{Dehghanian2020} discussed the idea of a DW in an attempt to
explain the 2014 HST observations of NGC~5548 (the AGN STORM~1
campaign). In this case, the normally correlated UV continuum and
broad emission-line variations de-coupled \citep{Goad2016,
  Pei2017}. The wind was suggested to rise almost perpendicularly to
the disk's surface, on the line connecting the ionizing source to the
BLR, and its base was assumed to be optically thick to some but not
all of the ionizing radiation. This changes considerably the level of
ionization and the emergent BLR spectrum. It also contributed broad,
high-velocity components to the high ionization lines. As the wind
rises from the disk's surface, part of it becomes observable through
X-ray and UV broad absorption features. The wind became more tenuous
with time, diminishing absorption features over several weeks.

A similar spectral behavior was observed in Mrk~817 and has been
discussed in previous papers of this series. Two scenarios were
proposed to explain the observed changes in the overall geometry of
the source, one based on the changing lags of the \civ\ line and one
on the changing continuum lags.  \cite{Homayouni2024} divided the
first 201 days of the campaign into three temporal windows, spanning
50, 55, and 96 days. They then correlated the 1144\AA\ continuum with
the \civ\ line and discovered very different lags of 11.7, 1.9, and
3.9 days, respectively.  They proposed a diminishing obscuration of
the BLR by a shielding DW, which resulted in a considerable shortening
of the lag of the \civ\ line. There is clear evidence, from UV and
X-ray observations, for the presence of wind along our line-of-sight
to the source in the first part of this period.  The analysis was
carried out on the total line flux and did not consider the
substantial lag differences between the line core and its red wing
found in \cite{Homayouni2023}.

L24 investigated a scenario of a DW that changes its direction and
column density over 420 days starting from the beginning of the
campaign. According to L24, this resulted in large observed
differences in continuum lags between Epoch~1 (HJD=2459177-2459317)
that overlap with temporal windows\,1 and 2 and part of temporal
window\,3, Epoch~2 (HJD=2459317-2459457), and Epoch~3
(HJD=2459457-2459597).  Possible geometrical configurations leading to
the two scenarios are illustrated schematically in
\cite{Homayouni2024} and L24.

 The idea of an obscuring DW during the first 140 days of the campaign
 is not without difficulties.  A detailed look into the \civ\ line
 behavior (Fig.~8 of \cite{Homayouni2023}) shows that the longest
 measured lags are associated mostly with the line core and the
 shortest ones with the high-velocity line wings, as expected from a
 gravitationally bound system of clouds. Given the unusual driving
 light curve of Mrk~817 discussed in section~\ref{sec:simulations} and
 the short duration of the three temporal windows, the impression of
 large changes in the lag of the {\it entire line} may be due to
 different ways of combining the core and the wings of the line rather
 than a large column density wind that obscures the BLR from the
 ionizing radiation. Moreover, such a DW during Epoch~1 would result
 in intense DC emission and considerably shorter continuum lags unless
 a very special geometry is assumed. Such shortening is not observed.

\begin{figure} \centering
        \includegraphics[width=0.99\linewidth]{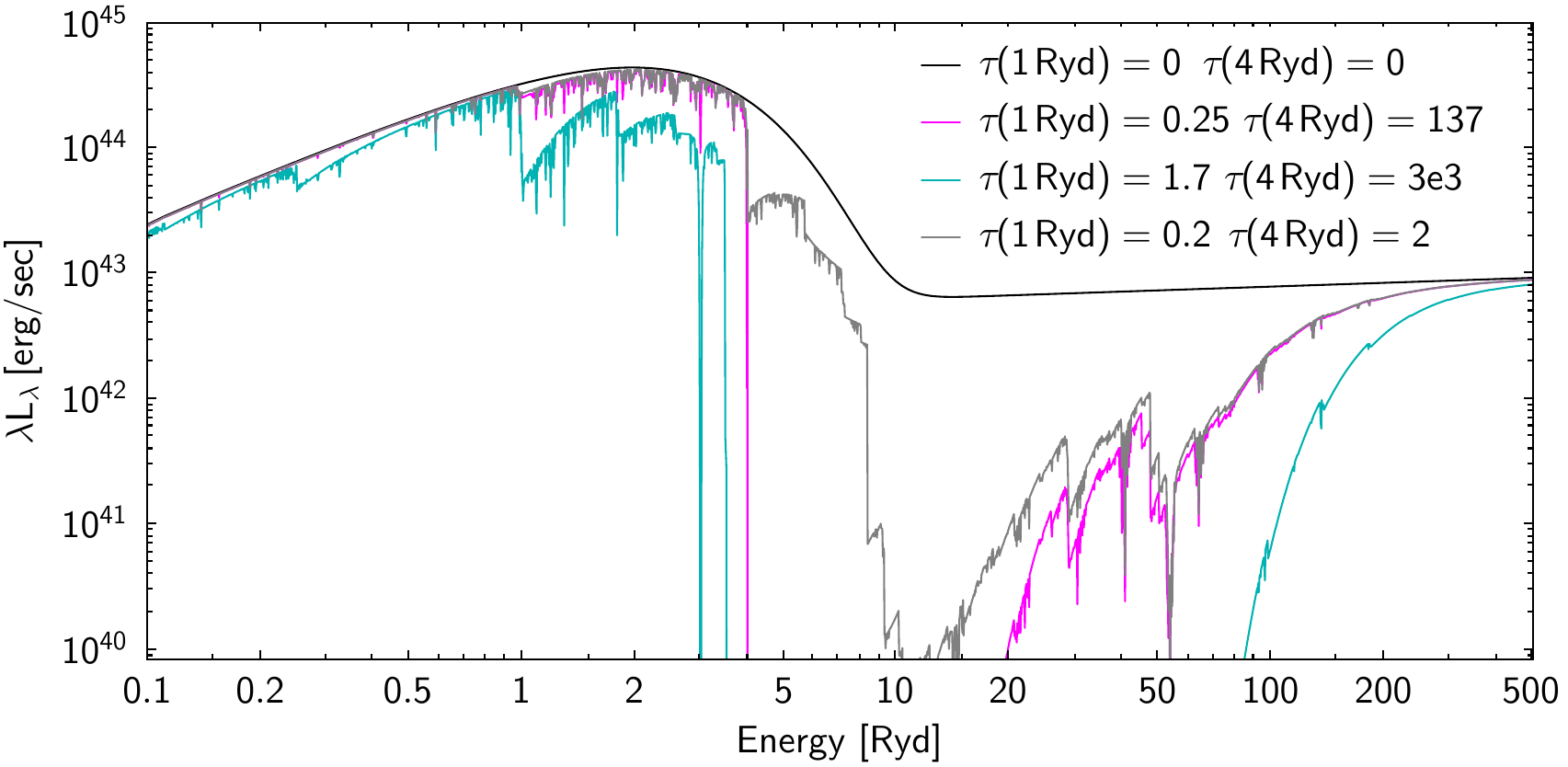}
	\caption{Various absorbed SEDs characterized by their optical depths at 1 and 4 Ryd.
}
 \label{transmitted_SEDs}
\end{figure}
          
\begin{table}

\centering
	\caption{Fractional transmitted luminosity.}
\begin{tabular}{lcccc}
\hline
SED & $\tau$(1~Ryd) & $\tau$(4~Ryd)  & 1-100~Ryd  & 4-100~Ryd \\ 
\hline
1 (black)   &  0          &  0           &  1.0       & 0.146$^a$   \\
2 (grey)   & 0.2        &  2.0         &  0.769     & 0.029    \\
3 (magenta)  & 0.25        &  137         &  0.744     & $<10^{-3}$ \\
4 (cyan)  & 1.7        & $>10^3$      &  0.335     & $<10^{-3}$\\ 
\hline
\end{tabular}
  \label{tab:fraction}
  $^a$L(1-100~Ryd)=6.23$\times 10^{44}$ ~erg/sec.
  
\end{table}

\subsection{Specific disk winds}
\label{sec:specific_DWs}
Below, we investigate four generic DW scenarios.  The properties of
the winds are defined by their location, column density, covering
factor, and optical depth as a function of time.  The scenarios are
distinguished by the flux transmitted through the wind material and
are illustrated in Fig.~\ref{transmitted_SEDs} and
Table~\ref{tab:fraction}. We explored a range of wind locations
between two and four light days from the BH.  The conclusions about
the line and continuum luminosities for shielded BLRs depend only
weakly on this choice. Still, line velocity and variability from the
gas in the wind depend on this assumption. The four scenarios are:
\begin{enumerate}
\item
 All ionizing luminosity is absorbed by a high column density wind
 (not shown in Fig.~\ref{transmitted_SEDs}). In this case, the wind
 can be considered a collection of RPC-BLR clouds similar to the ones
 assumed in Model~1 but probably at a shorter distance from the
 central black hole.

\item
In this case, represented by a grey line, the fraction of the
4-100~Ryd radiation absorbed by the wind is relatively large, and the
absorbed fraction of the 1-4~Ryd continuum is negligible.  For
equatorial winds of this type, the effect of obscuration is minor
spectral variations in several strong UV emission lines but no DC
emission variations.

 \item
The wind absorbs all the 4-100~Ryd flux (magenta line in
Fig.~\ref{transmitted_SEDs}) but only a small fraction of the 1-4~Ryd
flux.  The total reduction in the 1-100~Ryd flux is small, of order
25\%.  For shielded BLR gas, small changes in high ionization lines
are expected.

\item
The wind absorbs all the 4-100~Ryd flux and $\approx 50$\% of the
1-4~Ryd flux (cyan line in Fig.~\ref{transmitted_SEDs}). For shielded
BLR gas, significant variations in line and DC emissions are expected.
This case is the one most similar to the ''equatorial obscurer''
investigated by \cite{Dehghanian2020}.

\end{enumerate}

\subsubsection{Case (i): RPC DWs}
\label{sec:optically_thick_DW}

Our model~1, shown in green in all previous figures, represents a
small-size BLR but could also be used to study the properties of a DW
with similar size and location. An outflow of such ``RPC Wind'' will
affect the total observed line and continuum luminosities and their
lags during parts of the campaign. The assumed geometrical covering
factor of the wind is 0.2-0.3.  In the example shown so far (Model~1),
$\log R{(\rm cm)}= 15.75-16$, but smaller distances have also been
considered (see Appendix A). As in all RPC models, the gas density is
determined by the local radiation pressure of the central
source. Close to the ionization front inside typical clumps, this is
10$^{12.4 - 12.9}$\cc.

The direction of the wind's motion and its density and column density
change over time. Far from the disk, the gas is so dilute that its
emission is practically unobserved.  We assume that close to the
launch locations, the direction is perpendicular to the disk's
surface, and the velocity is close to the escape velocity at this
location. Given the chosen inclination of the BLR disk (41.4 degrees),
we can calculate the observed wind velocity.  The gas in the wind can
block the ionizing radiation from reaching the entire BLR, part of the
BLR, or no part at all.  For the latter case, the total emission from
the system (DC$+$lines) will be enhanced.

The column density of the clumps in the wind is an important
consideration. For all models discussed so far, it was assumed to be
$10^{23.5}$\,cm$^{-2}$. At a distance of $10^{16}$~cm or larger, this
is consistent with gravitationally bound clouds. At smaller distances,
the column density of the ionized gas is larger (at the back of the
cloud $n_e \approx n_{HI}$), and a sizeable fraction of the ionization
radiation can escape from the far side of the clouds. We have
therefore experimented with thicker clouds with a column density of
$10^{24}$\,cm$^{-2}$ and show several examples in Appendix~C.

The lag of the DC radiation emitted by the gas in Case-$i$ wind is
much smaller than the lag of the BLR gas, primarily because of the
much closer location of this component
(Fig.~\ref{mean_emissivity}). The high ionization lines like \civ\ are
predicted to be very broad compared with their FWHM in the BLR, and
the lags of the very broad wings are very short. Such winds produce
very weak Balmer lines.

 Given this, we suggest that an ejection of a DW with the properties
 assumed here (Model~1) could occur close to the end of Epoch~1,
 provided the wind is rising rapidly and changes its 1-4~Ryd optical
 depth over a time which is short compared with 140 days (the duration
 of Epoch~1 and Epoch~2). In fact, following the observed Lyman and
 Balmer lines and the core of \civ\ during Epoch~1 and Epoch~2, we
 could not identify any period where the line intensities are reduced
 by large factors as expected for complete shielding of the BLR.
 Later, the optical depth of the rising wind decreases, making it more
 similar to the other types of winds discussed here. As explained
 below, a Case-$iv$ wind best fits the conditions observed in
 Epoch\,2.

The wind ejection process described here is likely to happen
continuously during most of the AGN STORM 2 campaign. Thus, the
line-of-sight X-ray absorbing gas observed in Epoch~1 can be the
remnants of an ejection event before the beginning of Epoch~1, and the
DW observed during Epoch~2 can lead to the next epoch of X-ray
absorbing gas between HJD=2459530 and 2459600
\citep[see][]{Partington2023}.

  Finally, a comment about X-ray illumination and the location of a
  high column density DW.  The self-gravity radius of a standard thin
  disk, consistent with the mass and accretion rate measured for
  Mrk~817, is about 1000\Rg. This is also the estimated launch
  location of the wind, based on measured continuum lags, and perhaps
  also the high-velocity wings of \lya\ and \civ. The DW in question
  is a barrier between the central X-ray source and the extended disk,
  and it is hard to imagine X-ray-heated parts five or six light days
  from the BH unless the corona extends to extreme heights above the
  disk surface. Thus, high column density DWs and X-ray-illuminated
  regions at large distances seem inconsistent.

\subsubsection{Cases (ii) and (iii): X-ray winds and small optical depth equatorial winds}
\label{sec:x_ray_wind}

Cases (ii) and (iii) are characterized by medium to very high
$\tau$(4~Ryd) and small $\tau$(1~Ryd). We investigated their
contributions to the Mrk~817 observations assuming two locations:
close to the disk's surface inside the BLR (''equatorial wind'') and
along the line of sight to the source (``X-ray wind''). We provide
detailed calculations for case (iii) and discuss, in more general
terms, various types of X-ray winds.

We calculated a combined wind-BLR model made up of two components: An
equatorial DW illuminated by the bare high luminosity SED considered
earlier and a BLR identical in its geometrical properties to Model~2
($\log R({\rm cm})=16-17.25$) illuminated by the continuum transmitted
through the wind. We compared this model to the unshielded Model~2
discussed earlier.

\begin{figure} \centering
        \includegraphics[width=0.95\linewidth]{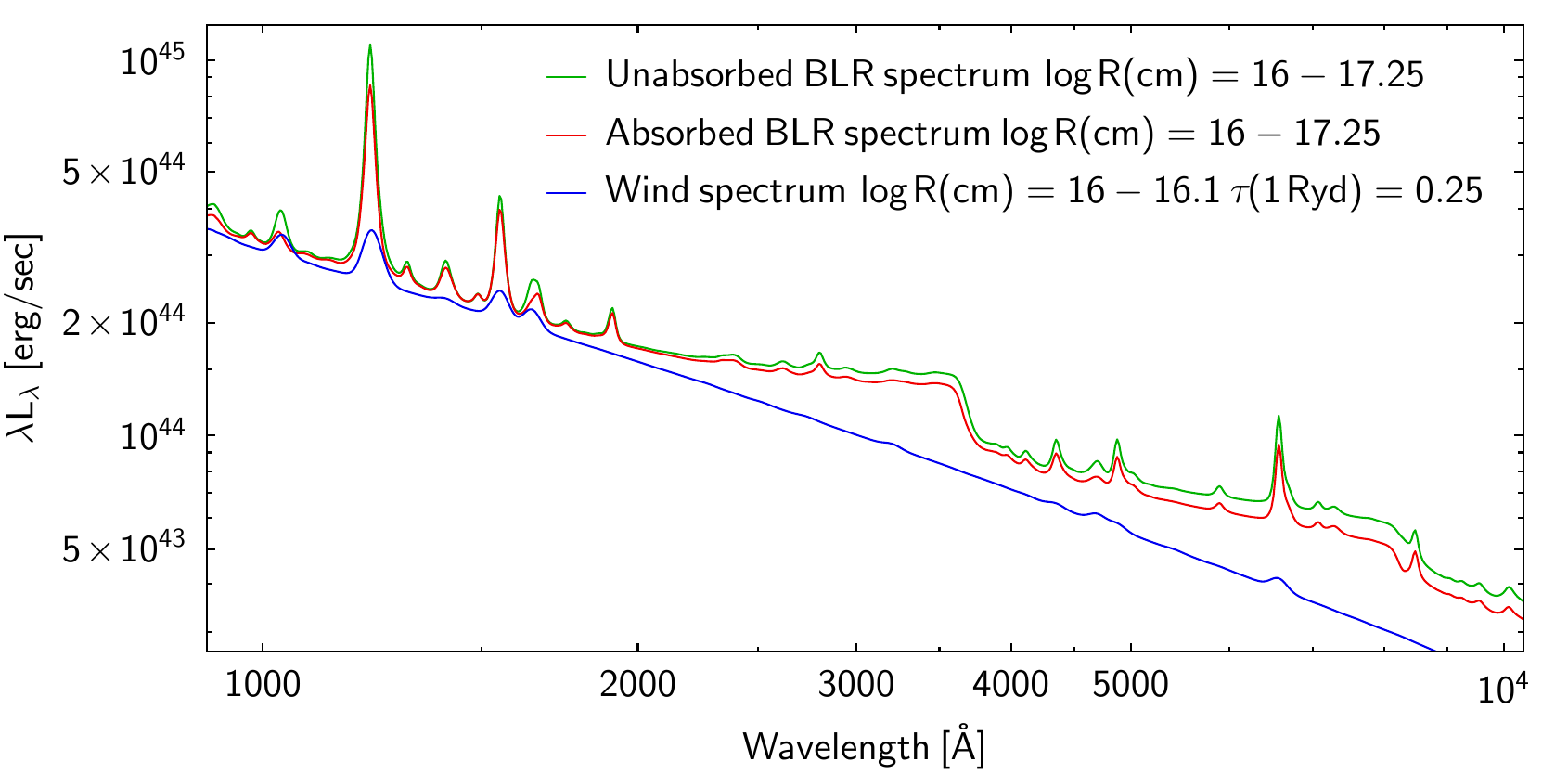}    
	\caption{Absorbed (red), unabsorbed (green), and Case (iii)
          wind (blue) spectra showing the significant contribution of
          the wind to the strong UV lines and an insignificant
          contribution to the DC emission.  }
 
 \label{wind_global_spectrum}
\end{figure}

Fig.~\ref{wind_global_spectrum} compares the global spectra of the
Case-(iii) wind, the shielded BLR, and the unshielded BLR.  Clearly,
for this wind, the broad wings of lines from ions of very high
ionization potential, like \heii$\lambda 1640$, are produced mostly by
the wind and lines with lower ionization energy, including \civ, have
contributions from both components.  The emission of hydrogen Lyman
and Balmer lines is dominated by the obscured BLR where conditions for
the production of these lines are more favorable.  The DC emission is
almost entirely from the shielded BLR.

Strong X-ray absorption features were observed over the entire Mrk~817
campaign. The NICER observations discussed in \cite{Partington2023}
provide estimates of the column density, line-of-sight covering
factor, and ionization parameter of this gas. The ionization parameter
is based on constant-density photoionization models. The similar
velocities of the broad \civ\ and \Siiv\ absorption lines with X-ray
absorption lines \citep{Zaidouni2024} during one epoch, combined with
the assumption that such velocities are of the order of the projected
escape velocity, provide a rough estimate of the location of the gas
at around $10^{16}$~cm. The observed spectrum is made of two
independent components, one marginally optically thick (X-ray wind)
and one with a very large optical depth (the BLR), at very different
distances.

We calculated the expected emission from clouds with properties
similar to the mean properties observed by \cite{Partington2023}
during Epoch~1: a column density of $10^{23}$~cm$^{-2}$, and an
ionization parameter of $\xi=L(1-1000~{\rm Ryd})/nR^2 \approx 1$.  We
assumed that the X-ray wind extends from 10$^{16}$ to 10$^{16.1}$~cm
(the exact range is not critical since the ionization parameter was
assumed to be the same in all clouds). While the global covering
factor during Epoch~1 is unknown, we can use the changing column
density and line-of-sight covering factors listed in
\cite{Partington2023} to estimate a mean (over time) global covering
factor of 0.2-0.4.

The calculated joint spectrum (not shown here) is similar to the
combined spectra shown in Figs.~\ref{wind_global_spectrum}. In
particular, an X-ray wind with the mean properties observed in Epoch~1
contributes very little to the total DC emission observed during our
campaign.

We also compared our X-ray absorber model with an RPC model at the
same mean distance and density at the illuminated face, identical to
the density in the constant density model. The critical issue is the
total optical depth, which needs to be very large to justify the RPC
assumption. We found that some of the clouds, those with a column
density exceeding $10^{23.2}$~cm$^{-2}$, with small ionization
parameters, are indeed pressure confined. Thus, some of the constant
density clouds with the properties listed in \cite{Partington2023} are
inconsistent with the main assumption made in the present paper. We
did not attempt to calculate all these models in detail.

Given this comparison, the X-ray absorbing gas observed in Epoch~1
cannot significantly affect the observed continuum lags because of its
weak DC emission. On the other hand, a fraction of the observed line
flux during this epoch, and perhaps during different times, can be
attributed to such gas.

\subsubsection{Case (iv): Intermediate optical depth equatorial winds}
\label{case_iv}

 This case is the most similar to the NGC~5548 model proposed by
 \cite{Dehghanian2020}.  The 1\,Ryd and 4\,Ryd optical depths (see
 Table~\ref{tab:fraction} and the cyan curve in
 Fig.~\ref{transmitted_SEDs}) are such that about half of the 1-4~Ryd
 ionizing continuum is absorbed by the wind. Our wind calculations
 extend over the range $10^{15.75}-10^{16}$\,cm and assume a column
 density of $10^{22.8}$\,cm$^{-2}$ for individual clumps. They confirm
 that this column density is large enough to produce significant DC
 emissions from the wind and to dominate the high-velocity wings of
 lines like \lya\ and \civ.  The BLR is exposed to the partly shielded
 ionizing flux. The high ionization lines are much weaker than in the
 case of no shielding, and the DC is strong but not as strong as the
 DC luminosity of the wind.

\begin{figure} \centering
        \includegraphics[width=0.95\linewidth]{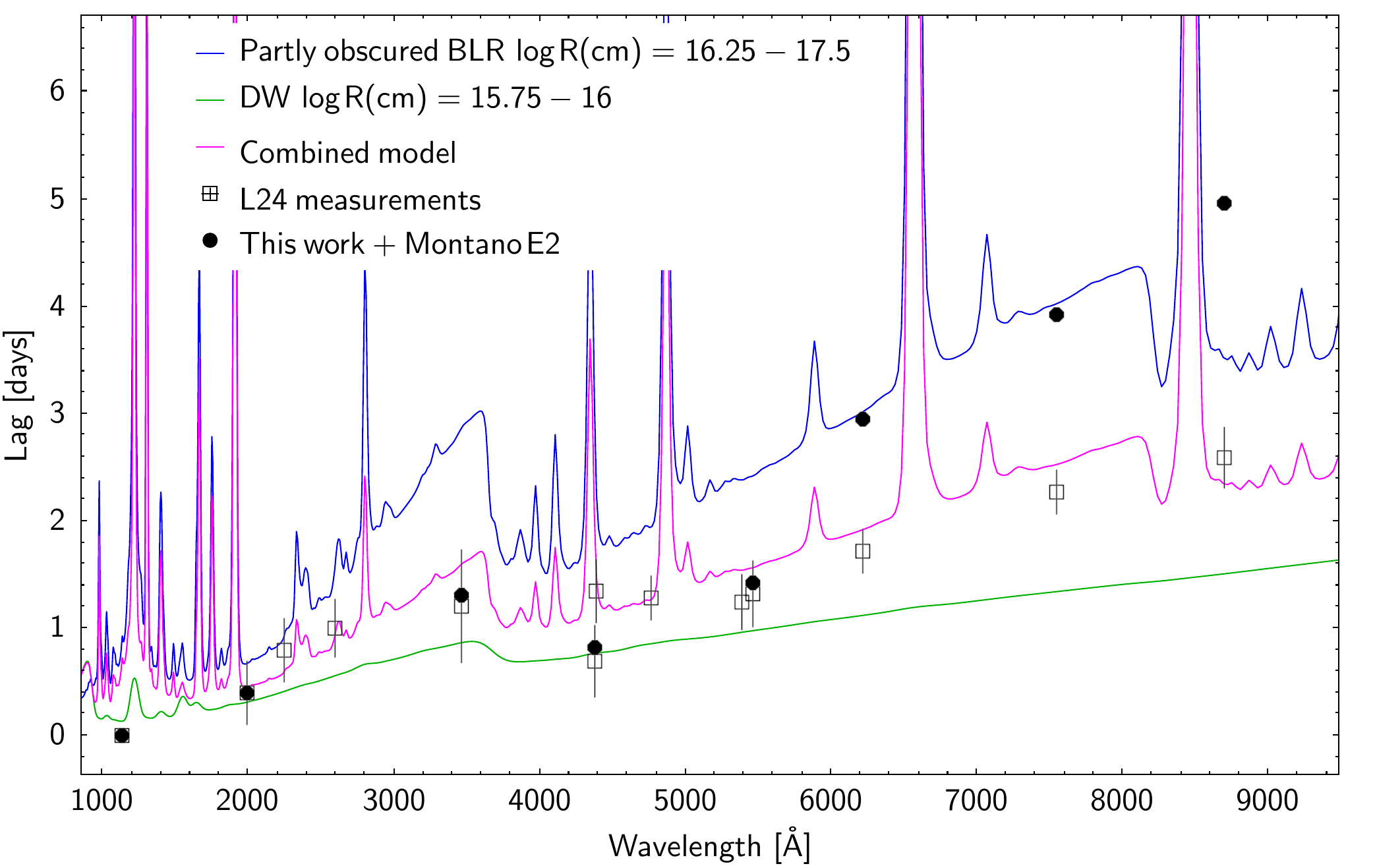}    
	\caption{The computed lags of a partly transparent disk wind
          (in green), partly shielded BLR (in blue), and their
          emissivity-weighted combination (magenta). The properties of
          the various components are marked in the figure. The models
          are compared with two ways of measuring lags in Epoch\,2:
          Frequency-resolved-based lags (L24) and ICCF-PyCCF lags
          (this work and Montano 2024). Note the sharp rise of the
          $r$, $i$, and $z$-band PyCCF-measured lags, which we
          interpret as due to the torus dust component (see text for
          more explanation).  }
\label{case_4_wind}
\end{figure}

Fig.~\ref{case_4_wind} shows three computed curves of
wavelength-dependent lags: One from the shielding wind (green line),
one from the partly obscured BLR (blue line), and the
emissivity-weighted lags for a combination of these two cases (magenta
line). Continuum RM measurements could easily detect the lag
differences between a pure BLR, for example, Model\,3 shown in
Fig.~\ref{observed_and_calculated_lags_model2_model3}, and a wind-BLR
combination like the one shown here.

Fig.~\ref{case_4_wind} also shows broad-band lags measured in two
different ways during Epoch~2: Frequency resolved lags from L24
computed for the 20-70\,${\rm day}^{-1}$ frequency range, and ICCF and
PyCCF lags calculated by us and by Montano et al. (in preparation).
The two sets of lags differ substantially at the longest wavelength
bands, suggesting that the $r$, $i$, and $z$-bands are affected by an
emission and/or reflection component at a distance of 70 or more light
days. As argued below, at least part of the difference can be
explained by dust emission from the nuclear torus in Mrk~817.

To summarize, the signature of a shielding disk wind, which absorbs
most of the ionizing radiation, is equivalent to that of a small
optically thick BLR at a very small distance from the BH, similar to
our Model~1.  This may result in a short transitory phase during which
the more distant BLR responds to the continuum flux emitted before the
wind is launched. Later, the line and continuum emission from the
obscured BLR declined dramatically, and most of the observed emission
is from the rising wind. There is no indication that this has happened
during the AGN STORM 2 campaign because all Balmer lines and other
strong emission lines do not show such behavior.

A rising large optical depth wind changes its properties and becomes
similar, for some time, to what was defined here as a Case-$iv$
wind. Such a phase could last 100-200 days and perhaps even
longer. During this time, the observed emission lines and DC emission
are contributed from both components: the rising wind and the partly
shielded BLR. As shown in Fig.~\ref{case_4_wind}, the calculated
continuum lags fit well the lags observed during Epoch~2, provided the
far-away dusty torus contribution is insignificant.

Cases $ii$ and $iii$ winds, those winds with large $\tau$(4~Ryd) and
small $\tau$(1~Ryd), represent well the X-ray absorbing winds observed
in Mrk~817.  Such winds result in minor variations of the spectrum of
the shielded BLR. In particular, the combined wind-shielded BLR model
is similar in spectral shape to the case of unshielded BLR, except
that the wings of the high ionization lines originating in the wind
component are broader and have shorter time lags.  There is no
significant DC emission by such DWs.

Our wind scenarios are meant to be general and do not follow all
possible time-dependent geometries. They are based on the 1 and 4~Ryd
optical depths and not on a systematic coverage of the density, column
density, and ionizing flux parameter space as in
\cite{Dehghanian2020}. Being sensitive to the changing optical depth
suggests that the properties of all those winds will change
considerably during the ``high-L'' and ``low-L'' episodes described
earlier.

\subsection{The dusty torus and long wavelength lags }

None of the models considered so far include thermal dust emission by
a torus-like structure beyond the sublimation radius of graphite
grains. Such structures are very common in high and low luminosity
AGN, and their properties have been discussed in numerous publications
\citep[e.g.,][and references therein]{Netzer2015}.  In particular,
near infra-red (NIR) monitoring of many sources
\citep[e.g.,][]{Koshida2014,GRAVITY2020b} show that the K-band NIR
flux lags the V-band continuum with a delay which is roughly 3-4 times
the typical delay of the \hb\ line. The observed K-band flux in
Mrk~817 suggests a similar torus with a delay of its innermost
illuminated face, made of pure graphite grains, of about 80
days. Given that much of the K-band flux is due to graphite grains
with temperatures exceeding about 1500K (the sublimation temperature
of silicate ISM grains), we should consider the effect of this
emission on the $i$ and $z$-bands of the source as well as the
additional delay relative to the optical-UV flux. This issue has been
discussed in several earlier publications, such as \cite{Honig2014},
\cite{Korista2019} and \cite{Netzer2022}.

A detailed investigation of dust-induced variations in Mrk\,817 is
beyond the scope of the present paper. Such an investigation should
consider various torus geometries, a range of dust temperatures, and,
most importantly, the nature of the optical-UV driving light
curve. Here, we only demonstrate the potential additional complexity
due to this process.

We consider the Case-$iv$ wind discussed earlier and the lag curves
shown in Fig.~\ref{case_4_wind}. We also assume that for Mrk~817,
$\lambda L_{\lambda} (2 \mu m) = 8.6 \times 10^{43}$\,\ergs\ (a
forthcoming AGN STORM 2 paper). We further consider several BBs with
T=1500-1700K and their contributions to the observed spectrum. The
case illustrated below assumes T(dust)=1600K.

\begin{figure} \centering
        \includegraphics[width=0.95\linewidth]{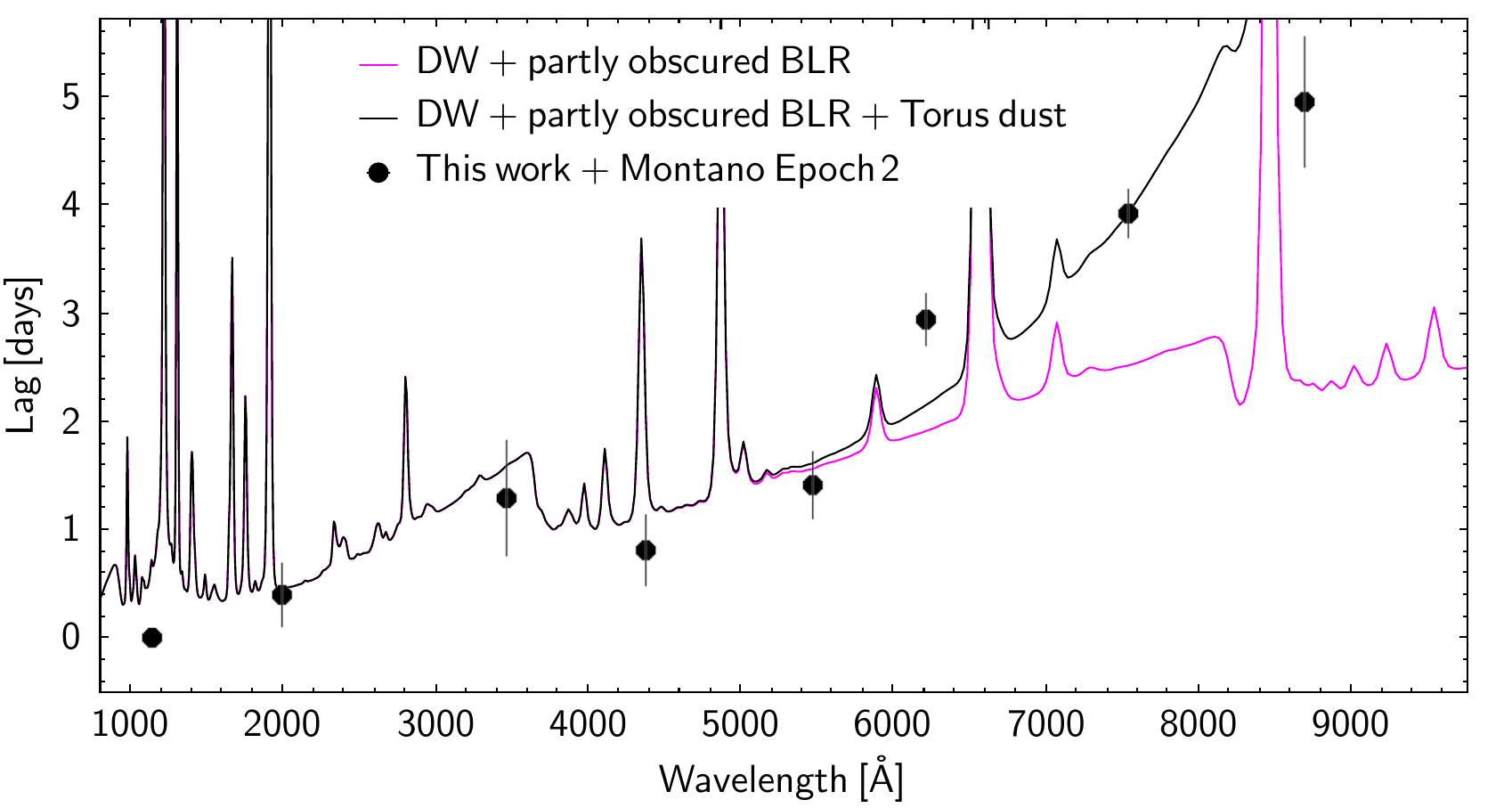}    
	\caption{The DW-BLR model from Fig.~\ref{case_4_wind} combined
          with torus dust assuming T(dust)=1600\,K.  }
 
\label{dust_delay}
\end{figure}

There are various ways to add the dust-induced lag to the illuminated
disk and DC-induced lags. The DC luminosity in our Case-$iv$ wind is
comparable to the accretion disk luminosity in the $i$ and
$z$-bands. Given this, we changed the definition of $L_{\rm tot}$ used
earlier to be $L_{\rm tot}=L_{\rm diff}+L_{\rm inc}+L_{\rm dust}$ and
added a new term, $\tau_{\lambda,{\rm dust}}$, defined as,
\begin{equation}
\tau_{\lambda,{\rm dust}} =80  f_{\rm lag}  \times 
   \frac{ L_{\rm dust} } { L_{\rm tot}}  \,\,{\rm days}  \,\,\, ,
\end{equation}
to Eqn.\,1. Here, 80 days is the assumed K-band delay based on the
mean measured \hb\ delay and $f_{\rm lag}=0.5$ as used in all BLR
models.  The results are shown by a black curve in
Fig.~\ref{dust_delay}.  This approximation shows how significant dust
delays can be at long wavelengths.

Our simplified estimate was not tested by numerical
simulations. Therefore, the value of $f_{\rm lag}$ is only a guess,
and the procedure of combining three variable components (disk,
diffuse emission, and dust) needs to be confirmed. Moreover, the
Epoch~2 period is only 140 days long, very short considering the
$\approx$ 80-day delay signature for the hot
dust. Fig.~\ref{tchar_tdur} in the appendix illustrates this issue. A
more likely possibility, which we cannot yet confirm, is a steady rise
or fall of $L_{\rm dust}$ during this period. Nevertheless, dust
delays cannot be excluded and should be more thoroughly investigated.

\subsection{Complications and uncertainties}

The models presented here focus on broad-band continuum lags with
uncertainties typical of this type of measurement. They are not meant
to provide a complete fit of the response and lags of all emission
lines. This would require a more careful analysis of the changes in
time and amplitude of the driving light curve with the location and
velocity-dependent response of the BLR gas.

Reproducing the observed interband continuum delays depends on various
assumptions and uncertainties.  Depending on redshift, all the
\swift\ and ground-based bands include one or more broad and variable
emission lines. The MERs of many lines differ from the MER of the DC
emission, which is the primary factor affecting the observed lags. An
important example is the \ciii$\lambda 1909$ emission line (see
Fig.~\ref{observed_and_calculated_lags_model2_model3} of this paper
and Fig.~13 in C23), which contaminates the \swift/UVW2
band. According to our Models\,2, and 3, the MER of this line is about
2-2.5 larger than the MER of the DC emission. This can increase the
lag in this band by 10-30\%.

A related issue is the computed intensity of the hydrogen Balmer
lines, the strong UV and optical \feii\ lines, and the \mgii\ doublet
at 2800\AA. As discussed in earlier publications, most recently in
\cite{Netzer2020, Netzer2022}, present BLR models tend to
underestimate the intensities and equivalent widths of these
lines. The too-weak computed UV \feii\ lines are evident in
Fig.~\ref{cosstis_model3_comparison}. The bands most affected by this
are the three \swift/UVOT bands and optical bands, which overlap with
the strong \ha\ and \hb\ lines. This tends to increase the observed
lags compared with the model predictions.

An uncertainty of a different type is the radial dependence of the
covering factor, $C(R)$. The power-law dependence assumed here is
common to several published BLR models. Its primary justification is
the good agreement between modeled and observed line lags. The simple
approximation we used, $dC(R) \propto R^{-\beta} dR$, with $\beta=2$,
is typical of successful BLR models, but somewhat different values of
$\beta$ are just as good. Additional possibilities not considered here
are non-radial distributions of optically thick and optically thin
clouds.

The assumed total covering factor of the clouds used in the
simulations (0.2) is likely smaller than the global geometrical
covering factor of the flared disk BLR. This would result in a
somewhat different TF but is unlikely to be a significant source of
uncertainty. Another uncertainty that depends on the BLR geometry and
the characteristics of the driving continuum is the exact value of
$f_{\rm lag}({\rm TF})$.  A factor not included in the present
calculations is the anisotropy of optically thick line emission,
particularly the Balmer lines.

Several other assumptions, such as the microturbulent velocity inside
the cloud and the gas metallicity, were not discussed in
detail. \cite{Netzer2020} discussed these in the general framework of
the RPC model and showed them to be of secondary importance if the
microturbulent velocity does not exceed about 30 \kms.

Finally, as Fig.~\ref{two_cosstis} shows, the disk SED in Mrk~817 is
changing with luminosity in a way that affects the ionizing SED, the
global level of ionization of the gas, and perhaps also the outer
boundary of the BLR. The present paper considers only a changing
luminosity not associated with a change in SED shape.  More advanced
accretion disk models are required to test other possibilities.

\section{Conclusions}

 We combined information from several papers in the AGN STORM~2 series
 about emission-line lags, X-ray and UV absorption features, and
 continuum lags with new BLR and disk-wind models to present a
 comprehensive view of the nucleus of Mrk~817.  Our main findings can
 be summarized as follows:
\begin{itemize}
\item
 Diffuse continuum (DC) emission, with additional contributions from
 strong and broad emission lines, can explain the broad-band continuum
 lags observed in this source. In all cases considered here, the
 X-ray-illuminated accretion disk contributes only a small fraction of
 the observed continuum lags, and there is no need to assume disk
 sizes exceeding the size of a standard thin accretion disk.
  \item 
  Our BLR models cover different possible geometries and a large range
  of distances from 2 to 122 light days. Modeling these assemblies of
  RPC clouds using the SED observed during the first 100 days of the
  campaign, we can explain the observed lags of \hb, \civ\, and
  several other emission lines. \cosstis\ spectra of the source during
  high and low flux levels indicate that the SED is bluer when
  brighter.  Computed models that assume luminosity variation but no
  change of SED shape are also consistent with the measured emission
  line and DC time lags.
  \item
  We suggest a modified procedure based on older methods and a new
  multiplication factor, $f_{\rm lag}({\rm TF})$, for combining DC
  lags with the lags predicted by simple, X-ray-illuminated thin
  accretion disks. We present extensive numerical simulations showing
  that $f_{\rm lag}({\rm TF}) \approx 0.5$ for typical BLR
  configurations. In more general situations, including a combination
  of disk wind and BLR, this factor depends on the specific TF and the
  nature of the driving continuum light curve.  We suggest that
  continuum light curves shorter than about 100 days may bias the
  measured lag in a way that questions the derived BLR geometry.
  \item 

  We suggest that Case-$iv$ winds, those with $\tau$(1\,Ryd )$\approx
  2$, were common in Mrk~817 during the AGN STORM~2 campaign. In
  particular, we show that the significant shortening of the continuum
  lags during the period HJD=2459317-2459457 (Epoch~2) is due to a
  combination of such a wind with a partly shielded BLR. This may well
  be a later stage of a larger column-density wind that was ejected
  from the disk at an earlier stage, at the end of Epoch~1. The wind
  does not obscure our line of sight to the source until much later,
  where its optical depth is further reduced (Case-$ii$ and Case-$iii$
  winds), and X-ray absorption features become evident.

  \item 
    BLR obscuration by small $\tau$(1~Ryd) winds leads to noticeable
    variations of the global observed spectrum, particularly the width
    and lags of several high ionization lines like \lya, \heii\ and
    \civ.

\item The launch locations of high column density DWs set upper limits
  on the distance of the parts of the disk that can be significantly
  heated by the central X-ray source.

 \end{itemize}

\begin{acknowledgments}

 This article is part of the series of papers by the AGN STORM 2
 collaboration. Our project began with the successful Cycle 28 HST
 proposal 16196 (Peterson et al. 2020). Support for Hubble Space
 Telescope program GO-16196 was provided by NASA through a grant from
 the Space Telescope Science Institute, which is operated by the
 Association of Universities for Research in Astronomy, Inc., under
 NASA contract NAS5-26555. We are grateful to the dedicated Institute
 staff who worked hard to review and implement this program. We
 particularly thank the Program Coordinator, W. Januszewski, who made
 sure the intensive monitoring schedule and coordination with other
 facilities continued successfully.

D.I., A.B.K, and L. \v C.P. acknowledge funding provided by the
University of Belgrade—Faculty of Mathematics (contract
451-03-66/2024-03/200104), Astronomical Observatory Belgrade (contract
451-03-66/2024- 03/200002), through grants by the Ministry of
Education, Science, and Technological Development of the Republic of
Serbia. D.I.  acknowledges the support of the Alexander von Humboldt
Foundation. A.B.K. and L. \v C.P. thank the support by Chinese Academy
of Sciences President’s International Fellowship Initiative (PIFI) for
visiting scientists.   NA acknowledges support from NSF grant AST
2106249, and NASA STScI grants AR-15786, AR-16600, AR-16601, and
HST-AR-17556.  G. Kriss and G. De Rosa acknowledge support from STScI
grant GO-16196.

We thank Jonathan Stern for useful discussions about RPC models. 

For the purpose of open access, the author has applied a Creative
Commons Attribution (CC BY) license to the Author Accepted Manuscript
version arising from this submission.

{\it Facilities:} \swift\ and HST (COS,STIS).

{\it Software:} $Cloudy$ \citep{Ferland2017}.

\end{acknowledgments}

%






\appendix

\section{Various mean-emissivity radii}

\begin{figure} \centering
        \includegraphics[width=0.95\linewidth]{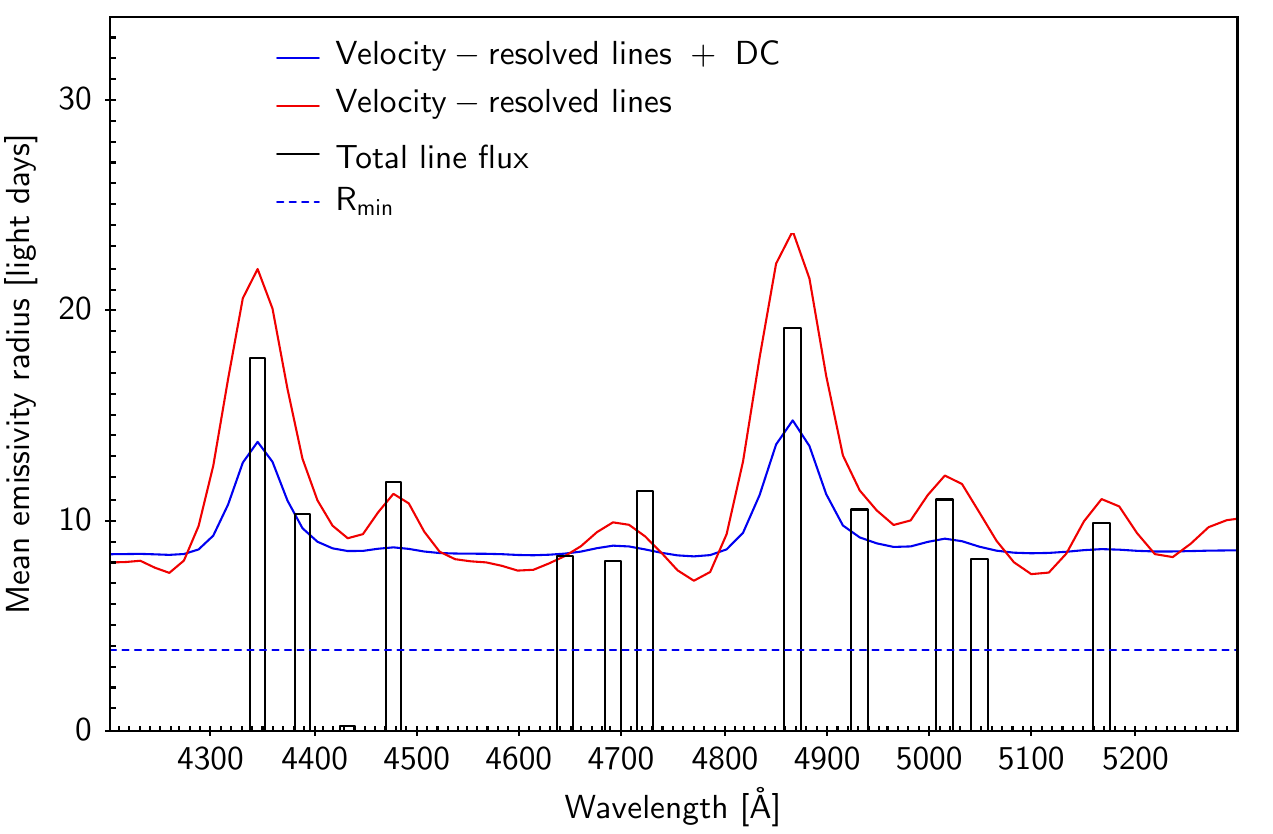}
	\caption{Various types of MERs over the \hg\ and \hb\ line
          region. Black: Integrated line luminosity MER (the quantity
          normally measured in 1D RM campaigns). Red:
          velocity-resolved line lags. Blue: broad-band MERs lags that
          include DC MERs in line-free regions of the spectrum or
          combinations of velocity-resolved lines and DC emission
          closer to the line core. This is normally smaller than the
          MER of velocity-resolved lines since the MER of the DC is
          considerably smaller ($\approx 50$\%) than the MER of the
          Balmer lines.
	}
\label{MER_near_hb}
\end{figure}

The exact meaning of the mean-emissivity radius (MER) used here
requires further explanation. For a broad emission line, the {\rm
  total} MER is the one usually compared with the lag of the total
line luminosity, which is integrated over a large velocity range and,
hence, for a gravitationally-bound BLR, a large range of
distances. Velocity-resolved RM MERs are smaller than the total MER in
the line wings and larger in the line core. A third type of MER is
shown in Fig.~\ref{mean_emissivity}. The vertical axis gives a
combined line$+$DC MER at each wavelength, which can differ from the
other MERs discussed here. An illustration of the three different MERs
in model~2 for the spectral region containing the \hb\ and \hg\ lines
is shown in Fig.~\ref{MER_near_hb}.

\section{Comparison with frequency-resolved lags}
\cite{Cackett2022}, and L24 used frequency resolved methods  \citep[e.g.,][]{Uttley2014}  to separate the total transfer function (or the impulse resolved function) into disk and BLR parts, 
\begin{equation}
 \psi_{\rm tot}(t) = (1-f_{\lambda})\psi_{\rm disk}(t) + f_{\lambda} \psi_{\rm BLR}(t),
 \label{impulse_response}
\end{equation}
where $f_{\lambda}$ represents the relative contributions of the BLR
to the total transfer function (TF) at wavelength ${\lambda}$ and
hence the physics and the size of the BLR. This approach avoids using
various time filters like detrending and Gaussian smoothing. They also
suggested representing $\psi_{BLR}$ by a two-parameter log-normal
distribution function, where the parameters $M$ and $S$ represent the
function's median ($e^M$) and standard deviation. L24 used the
observed \hb\ lag to suggest that the first part (Epoch 1) of the
campaign, when the obscuring X-ray column was large, can be
represented by a log-normal function with $e^M = 24$ and $S=0.9$
days. Epoch~2, when the X-ray column was much smaller, was described
by $e^M = 24$ days and $S=1.7$ days. Given this choice of parameters,
one can solve for $f_{\lambda}$ that best fits the observed lags.

\begin{figure} \centering
        \includegraphics[width=0.95\linewidth]{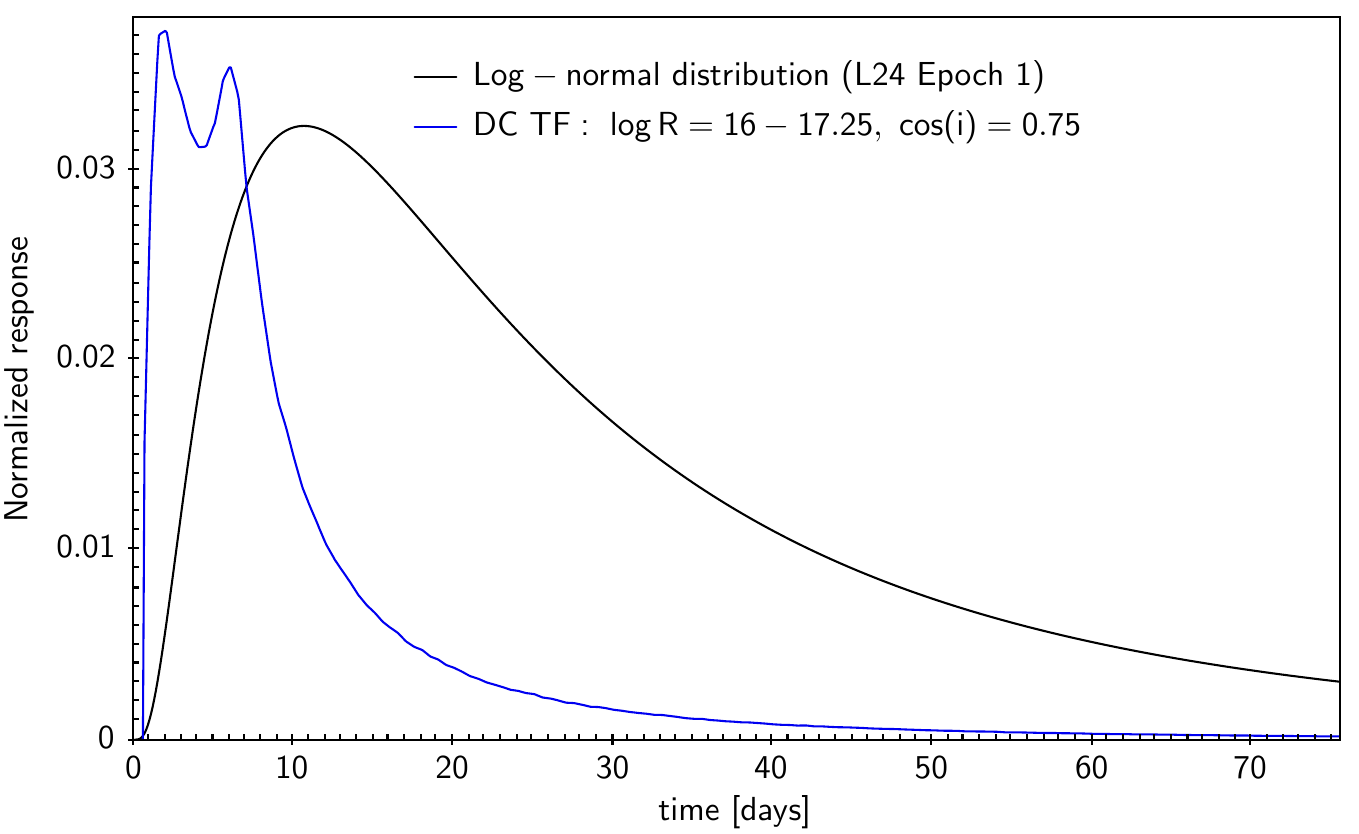}
	\caption{A comparison between the log-normal distribution used in L24 for Epoch 1 and the TF of the 7995\AA\ DC in our Model\,2 with $\log R({\rm cm})=16-17.25$. The mean-emissivity radii are 25 light days for the log-normal distribution and 11.3 light days for the 7995\AA\ TF.
    	 }
\label{appendix_1}
\end{figure}

 Our time-domain forward approach can be compared with the
 frequency-resolved analysis by comparing the transfer function of the
 DC (the dominant factor contributing to the continuum lags) to the
 \hb-based log-normal distribution function. This is shown in
 Fig.~\ref{appendix_1}. The two functions are very different, not so
 much because of their different shapes but mainly because the MER of
 the DC is much smaller ($\approx 50$\%) than the MER of the
 \hb\ line. This would result in very different contributions of the
 BLR to the total TF ($f_{\lambda}$ in eqn.~\ref{impulse_response}).


\section{Additional disk wind models}

Additional wind models that cover a range of distances equal to or
smaller than the range assumed in Model~1, all with a column density
of $10^{24}$cm$^{-2}$, have been computed.  In these cases, the
luminosity emitted over the Balmer continuum waveband is smaller than
the luminosity of the standard case shown in
Fig.~\ref{three_BLR_models} where the column density is
$10^{23.5}$\,cm$^{-2}$. The main reason is the increased opacity of
the gas over this band which results in more absorption of the
incident continuum. This increases the ionization of neutral hydrogen
and the locally emitted bound-free emission. However, much of the
additional emitted radiation spread over wavelength bands outside the
range shown here.

\begin{figure} \centering
        \includegraphics[width=0.95\linewidth]{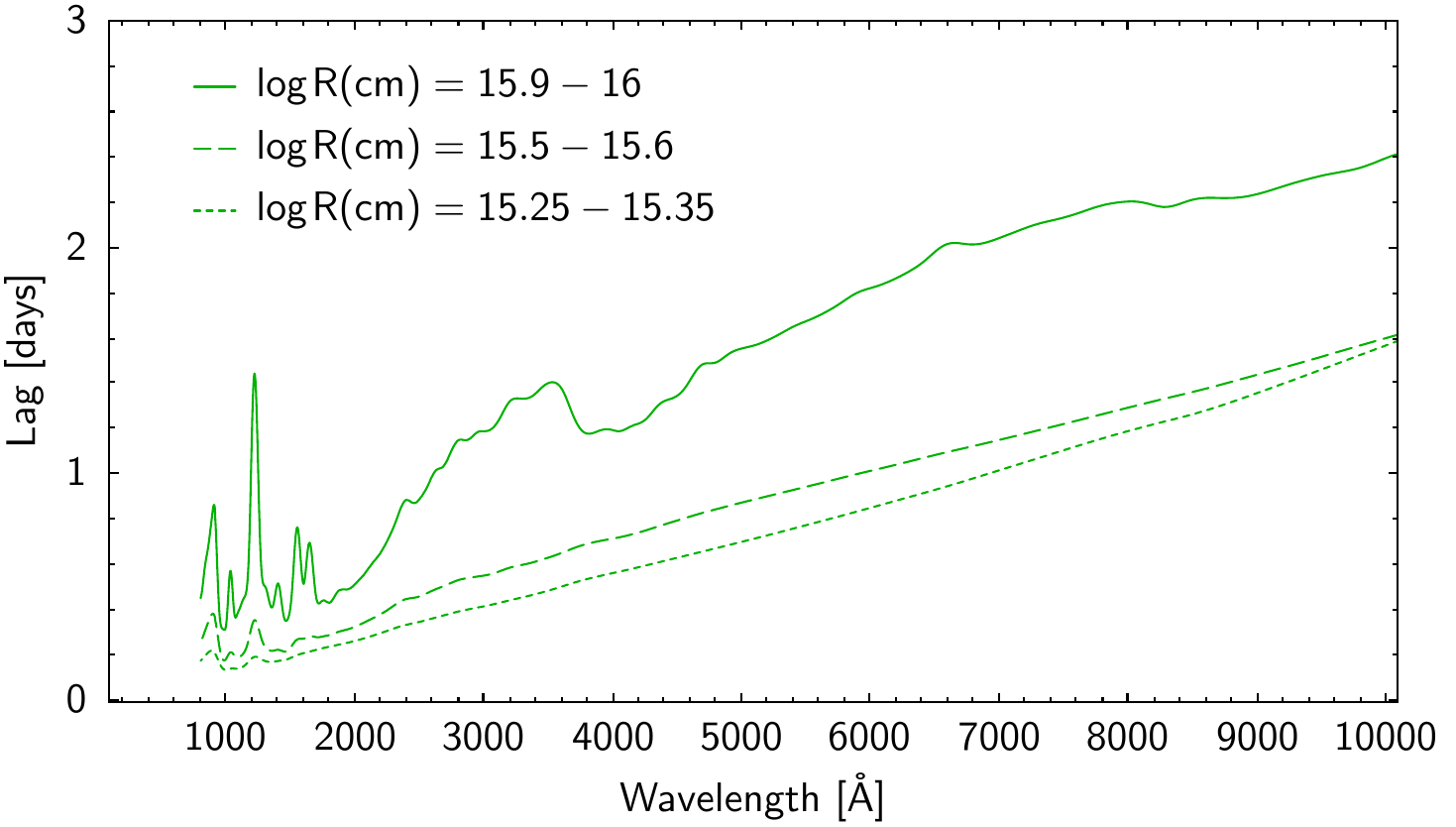}
        \caption{Wavelength dependent lags for winds with a column density of 10$^{24}$~cm$^{-2}$ and distances as marked.}  
        \label{appendix3}
\end{figure}

The computed lags for such winds are shown in
Fig.~\ref{appendix3}. Some lags are very short, similar to those
predicted for the X-ray illuminated disk.


\makeatletter
\relax

\section{Delay dependence on $T_{\rm char}$ and $T_{\rm dur}$}
\label{simulations}

\begin{figure}
    \centering
    \includegraphics[width=\columnwidth]{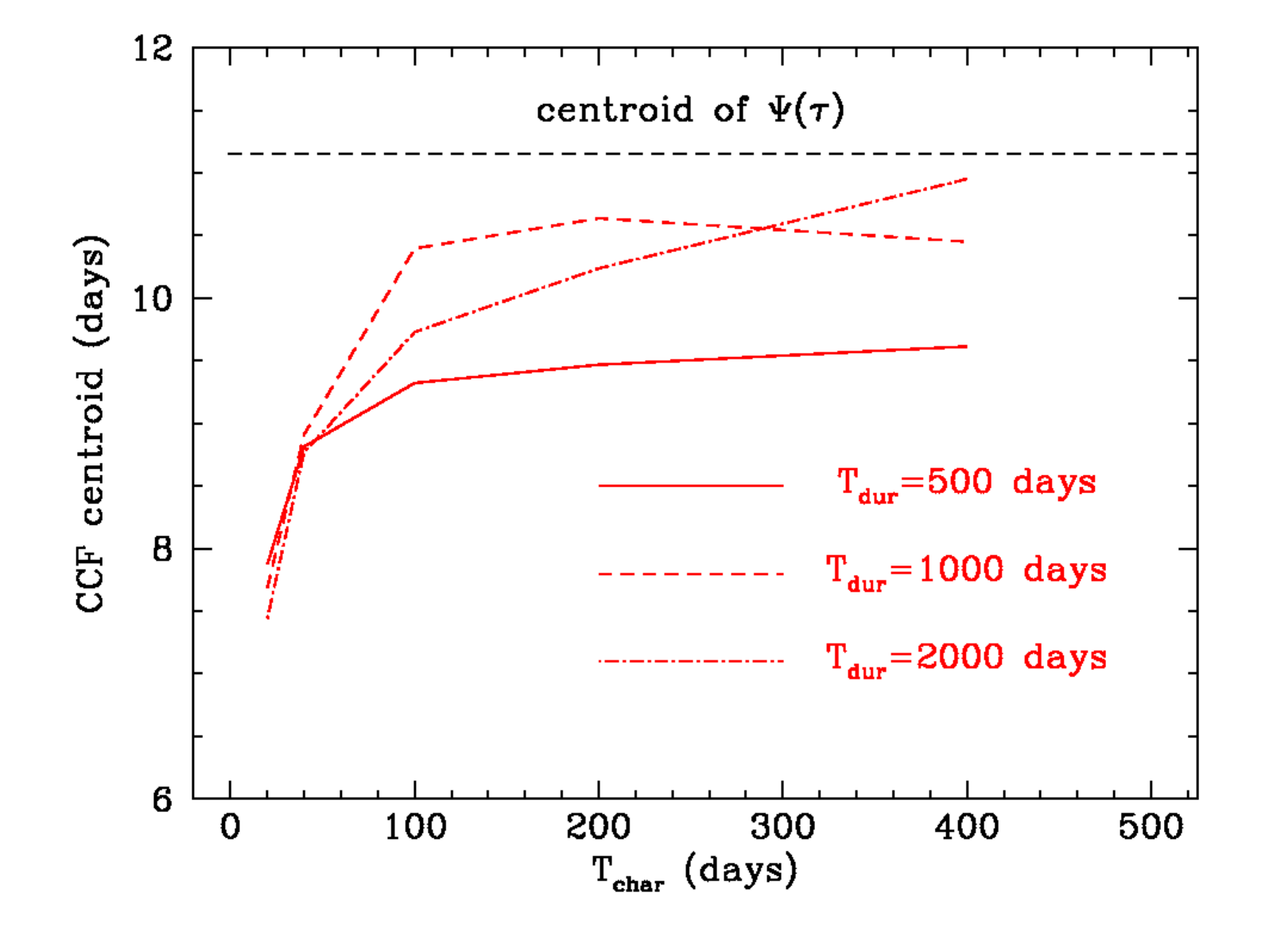}
    \caption{The measured lag (ICCF centroid) for the DC component of
      Model~2 relative to the driving continuum light-curve, here
      plotted as a function of the characteristic timescale of the
      driving continuum, for fixed light-curve durations of 500, 1000
      and 2000~days, each with 1~day sampling. In all cases, the
      measured ICCF centroid shows a strong dependence on $T_{\rm
        char}$, for small $T_{\rm char}$. Only for long light-curve
      durations does the ICCF centroid approach the MER, and then only
      for the largest $T_{\rm char}$.}
    \label{tchar_tdur}
\end{figure}

To illustrate the dependence of the measured lag of the DC component
on the characteristic timescale of the driving continuum $T_{\rm
  char}$, and campaign duration $T_{\rm dur}$, we drive our model BLR
(Model\,2) with simulated light-curves generated from a CAR1 process
(e.g., Kelly et~al. 2009), sampled at 1-day intervals. We fix the
variance of the simulated driving light curve but vary the damping
timescale, measuring the lag from the centroid of the
ICCF. Figure~\ref{tchar_tdur} illustrates the dependence of the
measured lag as a function of damping timescale, with damping
timescale in the range 20--400 days, and campaign durations of length
$T_{\rm dur}=500$, 1000 and 2000 days.

    In all cases, the measured delay shows a steep dependence on the
    damping timescale of the driving continuum for small $T_{\rm
      char}$, increasing as $T_{\rm char}$ increases. At large $T_{\rm
      char}$, this relation flattens. Only for large $T_{\rm char}$
    and long-duration campaigns does the measured delay approach the
    value expected from the measured centroid of the 1-d response
    function (indicated by the horizontal dashed line).  For the BLR
    model assumed here (Model~2), the number is very close to the
    lowest points of all three curves.
    



\bibliographystyle{aasjournal}
\bibliography{all_references_apr24}

\begin{thebibliography}{}
\expandafter\ifx\csname natexlab\endcsname\relax\def\natexlab#1{#1}\fi
\providecommand{\url}[1]{\href{#1}{#1}}
\providecommand{\dodoi}[1]{doi:~\href{http://doi.org/#1}{\nolinkurl{#1}}}
\providecommand{\doeprint}[1]{\href{http://ascl.net/#1}{\nolinkurl{http://ascl.net/#1}}}
\providecommand{\doarXiv}[1]{\href{https://arxiv.org/abs/#1}{\nolinkurl{https://arxiv.org/abs/#1}}}

\bibitem[{Baldwin {et~al.}(1995)Baldwin, Ferland, Korista, \&
  Verner}]{Baldwin1995}
Baldwin, J., Ferland, G., Korista, K., \& Verner, D. 1995, The Astrophysical
  Journal, 455, \dodoi{10.1086/309827}

\bibitem[{{Baskin} \& {Laor}(2018)}]{Baskin2018}
{Baskin}, A., \& {Laor}, A. 2018, \mnras, 474, 1970,
  \dodoi{10.1093/mnras/stx2850}

\bibitem[{{Baskin} {et~al.}(2014){Baskin}, {Laor}, \& {Stern}}]{Baskin2014}
{Baskin}, A., {Laor}, A., \& {Stern}, J. 2014, \mnras, 438, 604,
  \dodoi{10.1093/mnras/stt2230}

\bibitem[{{Bentz} {et~al.}(2009){Bentz}, {Peterson}, {Netzer}, {Pogge}, \&
  {Vestergaard}}]{Bentz2009}
{Bentz}, M.~C., {Peterson}, B.~M., {Netzer}, H., {Pogge}, R.~W., \&
  {Vestergaard}, M. 2009, \apj, 697, 160, \dodoi{10.1088/0004-637X/697/1/160}

\bibitem[{{Bentz} {et~al.}(2013){Bentz}, {Denney}, {Grier}, {Barth},
  {Peterson}, {Vestergaard}, {Bennert}, {Canalizo}, {De Rosa}, {Filippenko},
  {Gates}, {Greene}, {Li}, {Malkan}, {Pogge}, {Stern}, {Treu}, \&
  {Woo}}]{Bentz2013}
{Bentz}, M.~C., {Denney}, K.~D., {Grier}, C.~J., {et~al.} 2013, \apj, 767, 149,
  \dodoi{10.1088/0004-637X/767/2/149}

\bibitem[{{Bottorff} {et~al.}(2000){Bottorff}, {Korista}, \&
  {Shlosman}}]{Bottorff2000}
{Bottorff}, M.~C., {Korista}, K.~T., \& {Shlosman}, I. 2000, \apj, 537, 134,
  \dodoi{10.1086/309006}

\bibitem[{{Cackett} {et~al.}(2022){Cackett}, {Zoghbi}, \&
  {Ulrich}}]{Cackett2022}
{Cackett}, E.~M., {Zoghbi}, A., \& {Ulrich}, O. 2022, \apj, 925, 29,
  \dodoi{10.3847/1538-4357/ac3913}

\bibitem[{{Cackett} {et~al.}(2023){Cackett}, {Gelbord}, {Barth}, {De Rosa},
  {Edelson}, {Goad}, {Homayouni}, {Horne}, {Kara}, {Kriss}, {Korista}, {Landt},
  {Plesha}, {Arav}, {Bentz}, {Boizelle}, {Dalla Bont{\`a}}, {Dehghanian},
  {Donnan}, {Du}, {Ferland}, {Fian}, {Filippenko}, {Gonz{\'a}lez Buitrago},
  {Grier}, {Hall}, {Hu}, {Ili{\'c}}, {Kaastra}, {Kaspi}, {Kochanek},
  {Kova{\v{c}}evi{\'c}}, {Kynoch}, {Li}, {McLane}, {Mehdipour}, {Miller},
  {Montano}, {Netzer}, {Panagiotou}, {Partington}, {{\v{C}}. Popovi{\'c}},
  {Proga}, {Rogantini}, {Sanmartim}, {Siebert}, {Storchi-Bergmann},
  {Vestergaard}, {Wang}, {Waters}, \& {Zaidouni}}]{Cackett2023}
{Cackett}, E.~M., {Gelbord}, J., {Barth}, A.~J., {et~al.} 2023, \apj, 958, 195,
  \dodoi{10.3847/1538-4357/acfdac}

\bibitem[{Cai {et~al.}(2024)Cai, Wan, Cai, Fan, \& Wang}]{Cai2024}
Cai, M., Wan, Z., Cai, Z., Fan, L., \& Wang, J. 2024, Universe, 10,
  \dodoi{10.3390/universe10070282}

\bibitem[{{De Rosa} {et~al.}(2015){De Rosa}, {Peterson}, {Ely}, {Kriss},
  {Crenshaw}, {Horne}, {Korista}, {Netzer}, {Pogge}, {Ar{\'e}valo}, {Barth},
  {Bentz}, {Brandt}, {Breeveld}, {Brewer}, {Dalla Bont{\`a}}, {De
  Lorenzo-C{\'a}ceres}, {Denney}, {Dietrich}, {Edelson}, {Evans}, {Fausnaugh},
  {Gehrels}, {Gelbord}, {Goad}, {Grier}, {Grupe}, {Hall}, {Kaastra}, {Kelly},
  {Kennea}, {Kochanek}, {Lira}, {Mathur}, {McHardy}, {Nousek}, {Pancoast},
  {Papadakis}, {Pei}, {Schimoia}, {Siegel}, {Starkey}, {Treu}, {Uttley},
  {Vaughan}, {Vestergaard}, {Villforth}, {Yan}, {Young}, \& {Zu}}]{deRosa2015}
{De Rosa}, G., {Peterson}, B.~M., {Ely}, J., {et~al.} 2015, \apj, 806, 128,
  \dodoi{10.1088/0004-637X/806/1/128}

\bibitem[{Dehghanian {et~al.}(2020)Dehghanian, Ferland, Kriss, Peterson,
  Korista, Goad, Chatzikos, Guzm{\'{a}}n, Rosa, Mehdipour, Kaastra, Mathur,
  Vestergaard, Proga, Waters, Bentz, Bisogni, Brandt, Bont{\`{a}}, Fausnaugh,
  Gelbord, Horne, McHardy, Pogge, \& Starkey}]{Dehghanian2020}
Dehghanian, M., Ferland, G.~J., Kriss, G.~A., {et~al.} 2020, The Astrophysical
  Journal, 898, 141, \dodoi{10.3847/1538-4357/ab9cb2}

\bibitem[{Dehghanian {et~al.}(2024)Dehghanian, Arav, Kriss, Mehdipour, Byun,
  Walker, Sharma, Barth, Bentz, Boizelle, Brotherton, Cackett, Bonta, Rosa,
  Ferland, Fian, Filippenko, Gelbord, Goad, Horne, Homayouni, Ilic, Joner,
  Kara, Kaspi, Kochanek, Korista, Kosec, Kovacevic, Landt, Lewin, Partington,
  Popovic, Proga, Rogantini, Siebert, Storchi-Bergmann, Vestergaard, Waters,
  Wang, Zaidouni, \& Zu}]{Dehghanian2024}
Dehghanian, M., Arav, N., Kriss, G.~A., {et~al.} 2024, AGN STORM 2: VIII.
  Investigating the Narrow Absorption Lines in Mrk 817 Using HST-COS
  Observations.
\newblock \doarXiv{2407.04164}

\bibitem[{{Du} \& {Wang}(2019)}]{Du2019}
{Du}, P., \& {Wang}, J.-M. 2019, \apj, 886, 42,
  \dodoi{10.3847/1538-4357/ab4908}

\bibitem[{{Fausnaugh} {et~al.}(2016){Fausnaugh}, {Denney}, {Barth}, {Bentz},
  {Bottorff}, {Carini}, {Croxall}, {De Rosa}, {Goad}, {Horne}, {Joner},
  {Kaspi}, {Kim}, {Klimanov}, {Kochanek}, {Leonard}, {Netzer}, {Peterson},
  {Schn{\"u}lle}, {Sergeev}, {Vestergaard}, {Zheng}, {Zu}, {Anderson},
  {Ar{\'e}valo}, {Bazhaw}, {Borman}, {Boroson}, {Brandt}, {Breeveld}, {Brewer},
  {Cackett}, {Crenshaw}, {Dalla Bont{\`a}}, {De Lorenzo-C{\'a}ceres},
  {Dietrich}, {Edelson}, {Efimova}, {Ely}, {Evans}, {Filippenko}, {Flatland},
  {Gehrels}, {Geier}, {Gelbord}, {Gonzalez}, {Gorjian}, {Grier}, {Grupe},
  {Hall}, {Hicks}, {Horenstein}, {Hutchison}, {Im}, {Jensen}, {Jones},
  {Kaastra}, {Kelly}, {Kennea}, {Kim}, {Korista}, {Kriss}, {Lee}, {Lira},
  {MacInnis}, {Manne-Nicholas}, {Mathur}, {McHardy}, {Montouri}, {Musso},
  {Nazarov}, {Norris}, {Nousek}, {Okhmat}, {Pancoast}, {Papadakis}, {Parks},
  {Pei}, {Pogge}, {Pott}, {Rafter}, {Rix}, {Saylor}, {Schimoia}, {Siegel},
  {Spencer}, {Starkey}, {Sung}, {Teems}, {Treu}, {Turner}, {Uttley},
  {Villforth}, {Weiss}, {Woo}, {Yan}, \& {Young}}]{Fausnaugh2016}
{Fausnaugh}, M.~M., {Denney}, K.~D., {Barth}, A.~J., {et~al.} 2016, \apj, 821,
  56, \dodoi{10.3847/0004-637X/821/1/56}

\bibitem[{{Ferland} {et~al.}(2017){Ferland}, {Chatzikos}, {Guzm{\'a}n},
  {Lykins}, {van Hoof}, {Williams}, {Abel}, {Badnell}, {Keenan}, {Porter}, \&
  {Stancil}}]{Ferland2017}
{Ferland}, G.~J., {Chatzikos}, M., {Guzm{\'a}n}, F., {et~al.} 2017, \rmxaa, 53,
  385.
\newblock \doarXiv{1705.10877}

\bibitem[{{Goad} \& {Korista}(2015)}]{Goad2015}
{Goad}, M.~R., \& {Korista}, K.~T. 2015, \mnras, 453, 3662,
  \dodoi{10.1093/mnras/stv1861}

\bibitem[{Goad {et~al.}(1993)Goad, O'Brien, \& Gondhalekar}]{Goad1993}
Goad, M.~R., O'Brien, P.~T., \& Gondhalekar, P.~M. 1993, Monthly Notices of the
  Royal Astronomical Society, 263, 149, \dodoi{10.1093/mnras/263.1.149}

\bibitem[{{Goad} {et~al.}(2016){Goad}, {Korista}, {De Rosa}, {Kriss},
  {Edelson}, {Barth}, {Ferland}, {Kochanek}, {Netzer}, {Peterson}, {Bentz},
  {Bisogni}, {Crenshaw}, {Denney}, {Ely}, {Fausnaugh}, {Grier}, {Gupta},
  {Horne}, {Kaastra}, {Pancoast}, {Pei}, {Pogge}, {Skielboe}, {Starkey},
  {Vestergaard}, {Zu}, {Anderson}, {Ar{\'e}valo}, {Bazhaw}, {Borman},
  {Boroson}, {Bottorff}, {Brandt}, {Breeveld}, {Brewer}, {Cackett}, {Carini},
  {Croxall}, {Dalla Bont{\`a}}, {De Lorenzo-C{\'a}ceres}, {Dietrich},
  {Efimova}, {Evans}, {Filippenko}, {Flatland}, {Gehrels}, {Geier}, {Gelbord},
  {Gonzalez}, {Gorjian}, {Grupe}, {Hall}, {Hicks}, {Horenstein}, {Hutchison},
  {Im}, {Jensen}, {Joner}, {Jones}, {Kaspi}, {Kelly}, {Kennea}, {Kim}, {Kim},
  {Klimanov}, {Lee}, {Leonard}, {Lira}, {MacInnis}, {Manne-Nicholas}, {Mathur},
  {McHardy}, {Montouri}, {Musso}, {Nazarov}, {Norris}, {Nousek}, {Okhmat},
  {Papadakis}, {Parks}, {Pott}, {Rafter}, {Rix}, {Saylor}, {Schimoia},
  {Schn{\"u}lle}, {Sergeev}, {Siegel}, {Spencer}, {Sung}, {Teems}, {Treu},
  {Turner}, {Uttley}, {Villforth}, {Weiss}, {Woo}, {Yan}, {Young}, \&
  {Zheng}}]{Goad2016}
{Goad}, M.~R., {Korista}, K.~T., {De Rosa}, G., {et~al.} 2016, \apj, 824, 11,
  \dodoi{10.3847/0004-637X/824/1/11}

\bibitem[{{GRAVITY Collaboration} {et~al.}(2020){GRAVITY Collaboration},
  {Dexter}, {Shangguan}, {H{\"o}nig}, {Kishimoto}, {Lutz}, {Netzer}, {Davies},
  {Sturm}, {Pfuhl}, {Amorim}, {Baub{\"o}ck}, {Brandner}, {Cl{\'e}net}, {de
  Zeeuw}, {Eckart}, {Eisenhauer}, {F{\"o}rster Schreiber}, {Gao}, {Garcia},
  {Genzel}, {Gillessen}, {Gratadour}, {Jim{\'e}nez-Rosales}, {Lacour},
  {Millour}, {Ott}, {Paumard}, {Perraut}, {Perrin}, {Peterson}, {Petrucci},
  {Prieto}, {Rouan}, {Schartmann}, {Shimizu}, {Sternberg}, {Straub},
  {Straubmeier}, {Tacconi}, {Tristram}, {Vermot}, {Waisberg}, {Widmann}, \&
  {Woillez}}]{GRAVITY2020b}
{GRAVITY Collaboration}, {Dexter}, J., {Shangguan}, J., {et~al.} 2020, \aap,
  635, A92, \dodoi{10.1051/0004-6361/201936767}

\bibitem[{{Hern{\'a}ndez Santisteban} {et~al.}(2020){Hern{\'a}ndez
  Santisteban}, {Edelson}, {Horne}, {Gelbord}, {Barth}, {Cackett}, {Goad},
  {Netzer}, {Starkey}, {Uttley}, {Brandt}, {Korista}, {Lohfink}, {Onken},
  {Page}, {Siegel}, {Vestergaard}, {Bisogni}, {Breeveld}, {Cenko}, {Dalla
  Bont{\`a}}, {Evans}, {Ferland}, {Gonzalez-Buitrago}, {Grupe}, {Joner},
  {Kriss}, {LaPorte}, {Mathur}, {Marshall}, {Mehdipour}, {Mudd}, {Peterson},
  {Schmidt}, {Vaughan}, \& {Valenti}}]{Hernandez2020}
{Hern{\'a}ndez Santisteban}, J.~V., {Edelson}, R., {Horne}, K., {et~al.} 2020,
  \mnras, 498, 5399, \dodoi{10.1093/mnras/staa2365}

\bibitem[{{Homayouni} {et~al.}(2023){Homayouni}, {De Rosa}, {Plesha}, {Kriss},
  {Barth}, {Cackett}, {Horne}, {Kara}, {Landt}, {Arav}, {Boizelle}, {Bentz},
  {Brink}, {Brotherton}, {Chelouche}, {Dalla Bont{\`a}}, {Dehghanian}, {Du},
  {Ferland}, {Ferrarese}, {Fian}, {Filippenko}, {Fischer}, {Foley}, {Gelbord},
  {Goad}, {Gonz{\'a}lez Buitrago}, {Gorjian}, {Grier}, {Hall}, {Hern{\'a}ndez
  Santisteban}, {Hu}, {Ili{\'c}}, {Joner}, {Kaastra}, {Kaspi}, {Kochanek},
  {Korista}, {Kova{\v{c}}evi{\'c}}, {Kynoch}, {Li}, {McHardy}, {McLane},
  {Mehdipour}, {Miller}, {Mitchell}, {Montano}, {Netzer}, {Panagiotou},
  {Partington}, {Pogge}, {{\v{C}}. Popovi{\'c}}, {Proga}, {Rogantini},
  {Storchi-Bergmann}, {Sanmartim}, {Siebert}, {Treu}, {Vestergaard}, {Wang},
  {Ward}, {Waters}, {Williams}, {Zaidouni}, \& {Zu}}]{Homayouni2023}
{Homayouni}, Y., {De Rosa}, G., {Plesha}, R., {et~al.} 2023, \apj, 948, 85,
  \dodoi{10.3847/1538-4357/acc45a}

\bibitem[{{Homayouni} {et~al.}(2024){Homayouni}, {Kriss}, {De Rosa}, {Plesha},
  {Cackett}, {Goad}, {Korista}, {Horne}, {Fischer}, {Waters}, {Barth}, {Kara},
  {Landt}, {Arav}, {Boizelle}, {Bentz}, {Brotherton}, {Chelouche}, {Dalla
  Bont{\`a}}, {Dehghanian}, {Du}, {Ferland}, {Fian}, {Gelbord}, {Grier},
  {Hall}, {Hu}, {Ili{\'c}}, {Joner}, {Kaastra}, {Kaspi}, {Kova{\v{c}}evi{\'c}},
  {Kynoch}, {Li}, {Mehdipour}, {Miller}, {Mitchell}, {Montano}, {Netzer},
  {Neustadt}, {Partington}, {Popovi{\'c}}, {Proga}, {Storchi-Bergmann},
  {Sanmartim}, {Siebert}, {Treu}, {Vestergaard}, {Wang}, {Ward}, {Zaidouni}, \&
  {Zu}}]{Homayouni2024}
{Homayouni}, Y., {Kriss}, G.~A., {De Rosa}, G., {et~al.} 2024, \apj, 963, 123,
  \dodoi{10.3847/1538-4357/ad1be4}

\bibitem[{Hönig(2014)}]{Honig2014}
Hönig, S.~F. 2014, The Astrophysical Journal Letters, 784, L4,
  \dodoi{10.1088/2041-8205/784/1/L4}

\bibitem[{Ili{\'c} {et~al.}(2023)Ili{\'c}, Raki{\'c}, \&
  Popovi{\'c}}]{Ilic2023}
Ili{\'c}, D., Raki{\'c}, N., \& Popovi{\'c}, L.~C. 2023, The Astrophysical
  Journal Supplement Series, 267, 19, \dodoi{10.3847/1538-4365/acd783}

\bibitem[{{Kara} {et~al.}(2021){Kara}, {Mehdipour}, {Kriss}, {Cackett}, {Arav},
  {Barth}, {Byun}, {Brotherton}, {De Rosa}, {Gelbord}, {Hern{\'a}ndez
  Santisteban}, {Hu}, {Kaastra}, {Landt}, {Li}, {Miller}, {Montano},
  {Partington}, {Aceituno}, {Bai}, {Bao}, {Bentz}, {Brink}, {Chelouche},
  {Chen}, {Colmenero}, {Dalla Bont{\`a}}, {Dehghanian}, {Du}, {Edelson},
  {Ferland}, {Ferrarese}, {Fian}, {Filippenko}, {Fischer}, {Goad},
  {Gonz{\'a}lez Buitrago}, {Gorjian}, {Grier}, {Guo}, {Hall}, {Ho},
  {Homayouni}, {Horne}, {Ili{\'c}}, {Jiang}, {Joner}, {Kaspi}, {Kochanek},
  {Korista}, {Kynoch}, {Li}, {Liu}, {McHardy}, {McLane}, {Mitchell}, {Netzer},
  {Olson}, {Pogge}, {Popovi{\'c}}, {Proga}, {Storchi-Bergmann}, {Strasburger},
  {Treu}, {Vestergaard}, {Wang}, {Ward}, {Waters}, {Williams}, {Yang}, {Yao},
  {Zastrocky}, {Zhai}, \& {Zu}}]{Kara2021}
{Kara}, E., {Mehdipour}, M., {Kriss}, G.~A., {et~al.} 2021, \apj, 922, 151,
  \dodoi{10.3847/1538-4357/ac2159}

\bibitem[{Kaspi \& Netzer(1999)}]{Kaspi1999}
Kaspi, S., \& Netzer, H. 1999, The Astrophysical Journal, 524, 71,
  \dodoi{10.1086/307804}

\bibitem[{{Kaspi} {et~al.}(2000){Kaspi}, {Smith}, {Netzer}, {Maoz}, {Jannuzi},
  \& {Giveon}}]{Kaspi2000}
{Kaspi}, S., {Smith}, P.~S., {Netzer}, H., {et~al.} 2000, \apj, 533, 631,
  \dodoi{10.1086/308704}

\bibitem[{Korista {et~al.}(1997)Korista, Baldwin, Ferland, \&
  Verner}]{Korista1997}
Korista, K., Baldwin, J., Ferland, G., \& Verner, D. 1997, The Astrophysical
  Journal Supplement Series, 108, 401, \dodoi{10.1086/312966}

\bibitem[{{Korista} \& {Goad}(2000)}]{Korista2000}
{Korista}, K.~T., \& {Goad}, M.~R. 2000, \apj, 536, 284, \dodoi{10.1086/308930}

\bibitem[{{Korista} \& {Goad}(2019)}]{Korista2019}
---. 2019, \mnras, 489, 5284, \dodoi{10.1093/mnras/stz2330}

\bibitem[{{Korista} {et~al.}(1995){Korista}, {Alloin}, {Barr}, {Clavel},
  {Cohen}, {Crenshaw}, {Evans}, {Horne}, {Koratkar}, {Kriss}, {Krolik},
  {Malkan}, {Morris}, {Netzer}, {O'Brien}, {Peterson}, {Reichert},
  {Rodriguez-Pascual}, {Wamsteker}, {Anderson}, {Axon}, {Benitez}, {Berlind},
  {Bertram}, {Blackwell}, {Bochkarev}, {Boisson}, {Carini}, {Carrillo},
  {Carone}, {Cheng}, {Christensen}, {Chuvaev}, {Dietrich}, {Dokter},
  {Doroshenko}, {Dultzin-Hacyan}, {England}, {Espey}, {Filippenko}, {Gaskell},
  {Goad}, {Ho}, {Huchra}, {Jiang}, {Kaspi}, {Kollatschny}, {Laor}, {Luminet},
  {MacAlpine}, {MacKenty}, {Malkov}, {Maoz}, {Martin}, {Matheson}, {McCollum},
  {Merkulova}, {Metik}, {Mignoli}, {Miller}, {Pastoriza}, {Pelat}, {Penfold},
  {Perez}, {Perola}, {Persaud}, {Peters}, {Pitts}, {Pogge}, {Pronik}, {Pronik},
  {Ptak}, {Rawley}, {Recondo-Gonzalez}, {Rodriguez-Espinosa}, {Romanishin},
  {Sadun}, {Salamanca}, {Santos-Lleo}, {Sekiguchi}, {Sergeev}, {Shapovalova},
  {Shields}, {Shrader}, {Shull}, {Silbermann}, {Sitko}, {Skillman}, {Smith},
  {Smith}, {Snijders}, {Sparke}, {Stirpe}, {Stoner}, {Sun}, {Thiele}, {Tokarz},
  {Tsvetanov}, {Turnshek}, {Veilleux}, {Wagner}, {Wagner}, {Wanders}, {Wang},
  {Welsh}, {Weymann}, {White}, {Wilkes}, {Wills}, {Winge}, {Wu}, \&
  {Zou}}]{Korista1995}
{Korista}, K.~T., {Alloin}, D., {Barr}, P., {et~al.} 1995, \apjs, 97, 285,
  \dodoi{10.1086/192144}

\bibitem[{{Koshida} {et~al.}(2014){Koshida}, {Minezaki}, {Yoshii}, {Kobayashi},
  {Sakata}, {Sugawara}, {Enya}, {Suganuma}, {Tomita}, {Aoki}, \&
  {Peterson}}]{Koshida2014}
{Koshida}, S., {Minezaki}, T., {Yoshii}, Y., {et~al.} 2014, \apj, 788, 159,
  \dodoi{10.1088/0004-637X/788/2/159}

\bibitem[{Kova\v{c}evi\'{c} {et~al.}(2014)Kova\v{c}evi\'{c}, \v{C}.
  Popovi\'{c}, \& Kollatschny}]{Kovacevic2014}
Kova\v{c}evi\'{c}, J., \v{C}. Popovi\'{c}, L., \& Kollatschny, W. 2014,
  Advances in Space Research, 54, 1347,
  \dodoi{https://doi.org/10.1016/j.asr.2013.11.035}

\bibitem[{{Lawther} {et~al.}(2018){Lawther}, {Goad}, {Korista}, {Ulrich}, \&
  {Vestergaard}}]{Lawther2018}
{Lawther}, D., {Goad}, M.~R., {Korista}, K.~T., {Ulrich}, O., \& {Vestergaard},
  M. 2018, \mnras, 481, 533, \dodoi{10.1093/mnras/sty2242}

\bibitem[{{Lira} {et~al.}(2018){Lira}, {Kaspi}, {Netzer}, {Botti}, {Morrell},
  {Mej{\'{\i}}a-Restrepo}, {S{\'a}nchez-S{\'a}ez}, {Mart{\'{\i}}nez-Palomera},
  \& {L{\'o}pez}}]{Lira2018}
{Lira}, P., {Kaspi}, S., {Netzer}, H., {et~al.} 2018, \apj, 865, 56,
  \dodoi{10.3847/1538-4357/aada45}

\bibitem[{{Maoz} {et~al.}(1993){Maoz}, {Netzer}, {Peterson}, {Bechtold},
  {Bertram}, {Bochkarev}, {Carone}, {Dietrich}, {Filippenko}, {Kollatschny},
  {Korista}, {Shapovalova}, {Shields}, {Smith}, {Thiele}, \&
  {Wagner}}]{Maoz1993}
{Maoz}, D., {Netzer}, H., {Peterson}, B.~M., {et~al.} 1993, \apj, 404, 576,
  \dodoi{10.1086/172310}

\bibitem[{{Mej{\'\i}a-Restrepo} {et~al.}(2016){Mej{\'\i}a-Restrepo},
  {Trakhtenbrot}, {Lira}, {Netzer}, \& {Capellupo}}]{Mejia2016}
{Mej{\'\i}a-Restrepo}, J.~E., {Trakhtenbrot}, B., {Lira}, P., {Netzer}, H., \&
  {Capellupo}, D.~M. 2016, \mnras, 460, 187, \dodoi{10.1093/mnras/stw568}

\bibitem[{Naddaf {et~al.}(2023)Naddaf, Martinez-Aldama, Marziani, Panda,
  Sniegowska, \& Czerny}]{Naddaf2023}
Naddaf, M.~H., Martinez-Aldama, M.~L., Marziani, P., {et~al.} 2023, Astronomy
  \& Astrophysics, 675, A43, \dodoi{10.1051/0004-6361/202245698}

\bibitem[{{Netzer}(2013)}]{Netzer2013}
{Netzer}, H. 2013, {The Physics and Evolution of Active Galactic Nuclei}

\bibitem[{{Netzer}(2015)}]{Netzer2015}
---. 2015, \araa, 53, 365, \dodoi{10.1146/annurev-astro-082214-122302}

\bibitem[{Netzer(2020)}]{Netzer2020}
Netzer, H. 2020, Monthly Notices of the Royal Astronomical Society, 494, 1611,
  \dodoi{10.1093/mnras/staa767}

\bibitem[{{Netzer}(2022)}]{Netzer2022}
{Netzer}, H. 2022, \mnras, 509, 2637, \dodoi{10.1093/mnras/stab3133}

\bibitem[{{Neustadt} {et~al.}(2024){Neustadt}, {Kochanek}, {Montano},
  {Gelbord}, {Barth}, {De Rosa}, {Kriss}, {Cackett}, {Horne}, {Kara}, {Landt},
  {Netzer}, {Arav}, {Bentz}, {Dalla Bont{\`a}}, {Dehghanian}, {Du}, {Edelson},
  {Ferland}, {Fian}, {Fischer}, {Goad}, {Gonz{\'a}lez Buitrago}, {Gorjian},
  {Grier}, {Hall}, {Homayouni}, {Hu}, {Ili{\'c}}, {Joner}, {Kaastra}, {Kaspi},
  {Korista}, {Kova{\v{c}}evi{\'c}}, {Lewin}, {Li}, {McHardy}, {Mehdipour},
  {Miller}, {Panagiotou}, {Partington}, {Plesha}, {Pogge}, {Popovi{\'c}},
  {Proga}, {Storchi-Bergmann}, {Sanmartim}, {Siebert}, {Signorini},
  {Vestergaard}, {Zaidouni}, \& {Zu}}]{Neustadt2024}
{Neustadt}, J. M.~M., {Kochanek}, C.~S., {Montano}, J., {et~al.} 2024, \apj,
  961, 219, \dodoi{10.3847/1538-4357/ad1386}

\bibitem[{{O'Brien} {et~al.}(1994){O'Brien}, {Goad}, \&
  {Gondhalekar}}]{OBrien1994}
{O'Brien}, P.~T., {Goad}, M.~R., \& {Gondhalekar}, P.~M. 1994, \mnras, 268,
  845, \dodoi{10.1093/mnras/268.4.845}

\bibitem[{{Partington} {et~al.}(2023){Partington}, {Cackett}, {Kara}, {Kriss},
  {Barth}, {De Rosa}, {Homayouni}, {Horne}, {Landt}, {Zoghbi}, {Edelson},
  {Arav}, {Boizelle}, {Bentz}, {Brotherton}, {Byun}, {Dalla Bont{\`a}},
  {Dehghanian}, {Du}, {Fian}, {Filippenko}, {Gelbord}, {Goad}, {Gonz{\'a}lez
  Buitrago}, {Grier}, {Hall}, {Hu}, {Ili{\'c}}, {Joner}, {Kaspi}, {Kochanek},
  {Korista}, {Kova{\v{c}}evi{\'c}}, {Kynoch}, {McLane}, {Mehdipour}, {Miller},
  {Panagiotou}, {Plesha}, {Popovi{\'c}}, {Proga}, {Rogantini},
  {Storchi-Bergmann}, {Sanmartim}, {Siebert}, {Vestergaard}, {Ward}, {Waters},
  \& {Zaidouni}}]{Partington2023}
{Partington}, E.~R., {Cackett}, E.~M., {Kara}, E., {et~al.} 2023, \apj, 947, 2,
  \dodoi{10.3847/1538-4357/acbf44}

\bibitem[{{Pei} {et~al.}(2017){Pei}, {Fausnaugh}, {Barth}, {Peterson}, {Bentz},
  {De Rosa}, {Denney}, {Goad}, {Kochanek}, {Korista}, {Kriss}, {Pogge},
  {Bennert}, {Brotherton}, {Clubb}, {Dalla Bont{\`a}}, {Filippenko}, {Greene},
  {Grier}, {Vestergaard}, {Zheng}, {Adams}, {Beatty}, {Bigley}, {Brown},
  {Brown}, {Canalizo}, {Comerford}, {Coker}, {Corsini}, {Croft}, {Croxall},
  {Deason}, {Eracleous}, {Fox}, {Gates}, {Henderson}, {Holmbeck}, {Holoien},
  {Jensen}, {Johnson}, {Kelly}, {Kim}, {King}, {Lau}, {Li}, {Lochhaas}, {Ma},
  {Manne-Nicholas}, {Mauerhan}, {Malkan}, {McGurk}, {Morelli}, {Mosquera},
  {Mudd}, {Muller Sanchez}, {Nguyen}, {Ochner}, {Ou-Yang}, {Pancoast}, {Penny},
  {Pizzella}, {Poleski}, {Runnoe}, {Scott}, {Schimoia}, {Shappee}, {Shivvers},
  {Simonian}, {Siviero}, {Somers}, {Stevens}, {Strauss}, {Tayar}, {Tejos},
  {Treu}, {Van Saders}, {Vican}, {Villanueva}, {Yuk}, {Zakamska}, {Zhu},
  {Anderson}, {Ar{\'e}valo}, {Bazhaw}, {Bisogni}, {Borman}, {Bottorff},
  {Brandt}, {Breeveld}, {Cackett}, {Carini}, {Crenshaw}, {De
  Lorenzo-C{\'a}ceres}, {Dietrich}, {Edelson}, {Efimova}, {Ely}, {Evans},
  {Ferland}, {Flatland}, {Gehrels}, {Geier}, {Gelbord}, {Grupe}, {Gupta},
  {Hall}, {Hicks}, {Horenstein}, {Horne}, {Hutchison}, {Im}, {Joner}, {Jones},
  {Kaastra}, {Kaspi}, {Kelly}, {Kennea}, {Kim}, {Kim}, {Klimanov}, {Lee},
  {Leonard}, {Lira}, {MacInnis}, {Mathur}, {McHardy}, {Montouri}, {Musso},
  {Nazarov}, {Netzer}, {Norris}, {Nousek}, {Okhmat}, {Papadakis}, {Parks},
  {Pott}, {Rafter}, {Rix}, {Saylor}, {Schn{\"u}lle}, {Sergeev}, {Siegel},
  {Skielboe}, {Spencer}, {Starkey}, {Sung}, {Teems}, {Turner}, {Uttley},
  {Villforth}, {Weiss}, {Woo}, {Yan}, {Young}, \& {Zu}}]{Pei2017}
{Pei}, L., {Fausnaugh}, M.~M., {Barth}, A.~J., {et~al.} 2017, \apj, 837, 131,
  \dodoi{10.3847/1538-4357/aa5eb1}

\bibitem[{Peterson(2007)}]{Peterson2007}
Peterson, B.~M. 2007, 10.
\newblock \doarXiv{0703197}

\bibitem[{{Popovi{\'c}} {et~al.}(2019){Popovi{\'c}},
  {Kova{\v{c}}evi{\'c}-Doj{\v{c}}inovi{\'c}}, \&
  {Mar{\v{c}}eta-Mandi{\'c}}}]{Popovic2019}
{Popovi{\'c}}, L.~{\v{C}}., {Kova{\v{c}}evi{\'c}-Doj{\v{c}}inovi{\'c}}, J., \&
  {Mar{\v{c}}eta-Mandi{\'c}}, S. 2019, \mnras, 484, 3180,
  \dodoi{10.1093/mnras/stz157}

\bibitem[{Rees {et~al.}(1989)Rees, Netzer, \& Ferland}]{Rees1989}
Rees, M.~J., Netzer, H., \& Ferland, G.~J. 1989, The Astrophysical Journal,
  347, 640, \dodoi{10.1086/168155}

\bibitem[{Rosborough {et~al.}(2024)Rosborough, Robinson, Almeyda, \&
  Noll}]{Rosborough2024}
Rosborough, S.~A., Robinson, A., Almeyda, T., \& Noll, M. 2024, The
  Astrophysical Journal, 965, 35, \dodoi{10.3847/1538-4357/ad26f3}

\bibitem[{{Shen} {et~al.}(2016){Shen}, {Horne}, {Grier}, {Peterson}, {Denney},
  {Trump}, {Sun}, {Brandt}, {Kochanek}, {Dawson}, {Green}, {Greene}, {Hall},
  {Ho}, {Jiang}, {Kinemuchi}, {McGreer}, {Petitjean}, {Richards}, {Schneider},
  {Strauss}, {Tao}, {Wood-Vasey}, {Zu}, {Pan}, {Bizyaev}, {Ge}, {Oravetz}, \&
  {Simmons}}]{Shen2016}
{Shen}, Y., {Horne}, K., {Grier}, C.~J., {et~al.} 2016, \apj, 818, 30,
  \dodoi{10.3847/0004-637X/818/1/30}

\bibitem[{{Slone} \& {Netzer}(2012)}]{Slone2012}
{Slone}, O., \& {Netzer}, H. 2012, \mnras, 426, 656,
  \dodoi{10.1111/j.1365-2966.2012.21699.x}

\bibitem[{Stern {et~al.}(2014)Stern, Behar, Laor, Baskin, \&
  Holczer}]{Stern2014}
Stern, J., Behar, E., Laor, A., Baskin, A., \& Holczer, T. 2014, Monthly
  Notices of the Royal Astronomical Society, 445, 3011,
  \dodoi{10.1093/mnras/stu1960}

\bibitem[{Sun {et~al.}(2014)Sun, Wang, Chen, \& Zheng}]{Sun2014}
Sun, Y.-H., Wang, J.-X., Chen, X.-Y., \& Zheng, Z.-Y. 2014, The Astrophysical
  Journal, 792, 54, \dodoi{10.1088/0004-637X/792/1/54}

\bibitem[{{Uttley} {et~al.}(2014){Uttley}, {Cackett}, {Fabian}, {Kara}, \&
  {Wilkins}}]{Uttley2014}
{Uttley}, P., {Cackett}, E.~M., {Fabian}, A.~C., {Kara}, E., \& {Wilkins},
  D.~R. 2014, \aapr, 22, 72, \dodoi{10.1007/s00159-014-0072-0}

\bibitem[{{Vanden Berk} {et~al.}(2004){Vanden Berk}, {Wilhite}, {Kron},
  {Anderson}, {Brunner}, {Hall}, {Ivezi{\'c}}, {Richards}, {Schneider}, {York},
  {Brinkmann}, {Lamb}, {Nichol}, \& {Schlegel}}]{VandenBerk2004}
{Vanden Berk}, D.~E., {Wilhite}, B.~C., {Kron}, R.~G., {et~al.} 2004, \apj,
  601, 692, \dodoi{10.1086/380563}

\bibitem[{{Wamsteker} {et~al.}(1990){Wamsteker}, {Rodriguez-Pascual}, {Wills},
  {Netzer}, {Wills}, {Gilmozzi}, {Barylak}, {Talavera}, {Maoz}, {Barr}, \&
  {Heck}}]{Wamsteker1990}
{Wamsteker}, W., {Rodriguez-Pascual}, P., {Wills}, B.~J., {et~al.} 1990, \apj,
  354, 446, \dodoi{10.1086/168707}

\bibitem[{{Weaver} \& {Horne}(2022)}]{Weaver2022}
{Weaver}, J.~R., \& {Horne}, K. 2022, \mnras, 512, 899,
  \dodoi{10.1093/mnras/stac248}

\bibitem[{{Wills} {et~al.}(1985){Wills}, {Netzer}, \& {Wills}}]{Wills1985}
{Wills}, B.~J., {Netzer}, H., \& {Wills}, D. 1985, \apj, 288, 94,
  \dodoi{10.1086/162767}

\bibitem[{Zaidouni {et~al.}(2024)Zaidouni, Kara, Kosec, Mehdipour, Rogantini,
  Kriss, Behar, Kaastra, Barth, Cackett, Rosa, Homayouni, Horne, Landt, Arav,
  Bentz, Brotherton, Bontà, Dehghanian, Ferland, Fian, Gelbord, Goad,
  Buitrago, Grier, Hall, Hu, Ilić, Kaspi, Kochanek, Kovačević, Kynoch,
  Lewin, Montano, Netzer, Neustadt, Panagiotou, Partington, Plesha,
  Popovi{\'c}, Proga, Storchi-Bergmann, Sanmartim, Siebert, Signorini,
  Vestergaard, Waters, \& Zu}]{Zaidouni2024}
Zaidouni, F., Kara, E., Kosec, P., {et~al.} 2024, AGN STORM 2: IX. Studying the
  Dynamics of the Ionized Obscurer in Mrk 817 with High-resolution X-ray
  Spectroscopy.
\newblock \doarXiv{2406.17061}

\end{thebibliography}



\end{document}